\documentclass[fleqn,usenatbib]{mnras}

\usepackage{mathptmx}

\usepackage[T1]{fontenc}

\DeclareRobustCommand{\VAN}[3]{#2}

\let\VANthebibliography\thebibliography
\def\thebibliography{\DeclareRobustCommand{\VAN}[3]{##3}\VANthebibliography}

\usepackage{graphicx}	
\usepackage{amsmath}    
\usepackage{amssymb}	
\usepackage{subfigure}
\usepackage{xcolor}
\usepackage{bigints}
\usepackage{dashundergaps}
\usepackage{setspace,booktabs,multirow,supertabular,lscape,threeparttable}
\usepackage{float,colortbl}
\usepackage{placeins}
\usepackage{indentfirst}
\usepackage{enumitem}
\usepackage{setspace}
\usepackage{comment}
\usepackage{graphicx}
\usepackage{hyperref}

\newcommand{\passband}[1]{W_{#1}}

\newlist{mylist}{itemize}{4}
\setlist[mylist]{label*=\araBIC*., leftmargin=0pt, labelsep=1.5em, align=left, labelwidth=!}
\usepackage{float}

\newcommand{\rtcru}{RT\,Cru}
\newcommand{\chandra}{{\sl Chandra}}

\title[]{A novel approach to detect line emission under high background in high-resolution X-ray spectra}

\author[Zhang et al.]{
Xiangyu Zhang$^{1}$,
Sara Algeri$^{1*}$,
Vinay Kashyap$^{2*}$, and Margarita Karovska$^{2}$,
\\
$^{1}$School of Statistics, University of Minnesota, 224 Church St. SE, Minneapolis, MN, USA\\
$^{2}$Center for Astrophysics $|$ Harvard \& Smithsonian, 60 Garden St., Cambridge, MA 02138, USA\\
$^{*}$Corresponding authors. Emails: salgeri@umn.edu, vkashyap@cfa.harvard.edu.
}

\date{Accepted 2023 February 1. Received 2023 January 22; in original form 2022 December 3}

\pubyear{2023}

\begin{document}
\label{firstpage}
\pagerange{\pageref{firstpage}--\pageref{lastpage}}
\maketitle

\begin{abstract}
We develop a novel statistical approach to identify emission features or set upper limits in high-resolution spectra in the presence of high background.  The method relies on detecting differences from the background using smooth tests and using classical likelihood ratio tests to characterize known shapes like emission lines.  We perform signal detection or place upper limits on line fluxes while accounting for the problem of multiple comparisons.  We illustrate the method by applying it to a \chandra~LETGS+HRC-S observation of symbiotic star \rtcru, successfully detecting previously known features like the Fe line emission in the 6-7~keV range and the Iridium-edge due to the mirror coating on \chandra.  We search for thermal emission lines from Ne~X, Fe~XVII, O~VIII, and O~VII, but do not detect them, and place upper limits on their intensities consistent with a $\approx$1~keV plasma.  We serendipitously detect a line at 16.93~\AA\ (0.732 keV) that we attribute to photoionization or a reflection component.
\end{abstract}

\begin{keywords}
methods: data analysis – methods: statistical – binaries: symbiotic - stars: individual: RT Cru X-rays: stars techniques: spectroscopic.
\end{keywords}


\section{Introduction}
\label{s:intro}

Recently, several new X-ray missions and mission concepts have been developed with the goal of obtaining high-resolution spectra (e.g., XRISM [\citealt{2022arXiv220205399X,2022SPIE12181E..1SI}], Athena [\citealt{2022arXiv220814562B}], Arcus [\citealt{2020SPIE11444E..2CS}], LEM [\citealt{2022arXiv221109827K}]).  With such spectra expected to become a ubiquitous feature of high-energy investigations, it is necessary to consider the challenge of detecting weak features when the spectrum is also contaminated by high levels of background of both instrumental and cosmic origins.  Traditional methods employed for spectral analysis, like globally fitting simplified spectral models, or carrying out isolated analyses of individual spectral lines, will fail to produce robust results; the former when the data show more structure than is encoded in the models, and the latter when statistical fluctuations in the background confound the detection of weak features.

Here, we present a newly developed statistical approach to evaluate the significance of deviations that may be present in high-resolution spectra even in the presence of high background \citep{algeri2020,algeri2021}. 
The method works by comparing a postulated model for the background and a source-free background dataset,  to assess the validity of the former and,  if needed, provide a data-driven correction for it based on the information carried by the latter.  Once the background model has been  ``trained'' on the signal-free sample, our strategy is to employ \emph{smooth tests}, originally introduced by \citet{neyman1937}, to assess whether the source spectrum of interest shows any \emph{shape} differences from the re-calibrated background model.  Whenever significant deviations cannot be found, we then search for the presence of the most prominent spectral lines at well-defined locations by means of likelihood ratio tests and place strict upper limits on their fluxes following the framework of \citet{2010ApJ...719..900K}. The statistical properties of the procedure proposed, i.e., power and probability of type I error, are also investigated.

Our method is applicable to high-resolution spectra, i.e., one where the resolution of the detector is sufficient to separate individual emission lines.  In the case of low-resolution spectra, the presence or absence of individual lines must be inferred by modeling \citep[see, e.g.,][]{park}. In this setting, Bayesian solutions are often implemented in order to overcome the difficulties associated with low bin counts and the lack of model identifiability within the digitization limits of the spectra.  In contrast, here we implement an {\sl unbinned} analysis on data obtained as lists of photons. The goal is to detect differences between the likelihoods under the background-only and the background+source hypotheses.  

Estimation and inference can then be facilitated by conducting a separate analysis over different regions of the spectrum following a frequentist statistical paradigm. The inferential results are ultimately combined by implementing adequate corrections for multiple hypothesis testing. 

To illustrate the necessity of the method proposed here in the analysis of high-resolution spectra, we apply
it to study the symbiotic stellar system \rtcru\ through the analysis of the \chandra\ LETGS+HRC-S dataset. We emphasize that the method itself has been developed to have general applicability to high-energy high-resolution datasets where weak spectral features are expected over strong background contamination.
We describe the \rtcru~system and the \chandra\ data below in Section~\ref{s:data}, and demonstrate the need for a principled statistical method to obtain robust inferences.
We then provide a detailed description of \emph{smooth models} and \emph{smooth tests} in Section~\ref{s:methods}.
In Section~\ref{s:analysis}, we describe the analysis that we carry out. Specifically, in Section~\ref{sec:background_only} we focus on the implementation of data-driven background correction, in Section~\ref{sec:4.2} we implement the proposed testing procedure to detect spectral features, and in Section~\ref{sec:constructUL} we construct upper limits for specified line locations where no signal is detected.
The applicability of the method and the results of the analysis are discussed in Section~\ref{sec:discuss}.
We summarise our work in Section~\ref{sec:summary}.

\section{The relevance of \rtcru}
\label{sec:rtcru}

In order to demonstrate the suitability of smooth tests in detecting new signals when the data is contaminated by high background, we apply it to answer questions that arose during the analysis of \chandra\ spectra of \rtcru, and which could not be dealt with using existing methods.  Below, we describe the astronomical object, and the dataset, and illustrate how an improved statistical analysis is necessary.

\rtcru\ is a symbiotic system (distance 2.5~kpc; GAIA EDR3) where a high-mass white dwarf (WD; M$_{WD}>1.3$~M$_\odot$) accretes from the wind of an M5\,III red giant companion \citep{1994A&AS..106..243C}.  
Symbiotic systems have been invoked as potential progenitors of a fraction of SN type Ia - key cosmological indicators. 
Symbiotic systems, especially those containing a high-mass ($>$1.3~M$_{\odot}$) white dwarf component are important for understanding the possibility of a single degenerate path toward SN Ia.

Symbiotic systems produce soft X-ray spectra from the accretion disk surrounding the WD, through quasi-steady burning of the accreting material on the WD surface \citep{1997A&A...319..201M}, and from a jet. 

A handful of systems have also been detected in hard X-rays, including \rtcru\ (INTEGRAL, \citealt{2007ApJS..170..175B}; {\sl Swift}, \citealt{2009ApJ...701.1992K}; {\sl Suzaku}, \citealt{2016A&A...592A..58D}, \chandra, \citealt{2007ApJ...671..741L,2021MNRAS.500.4801D}; {\sl NuSTAR+Suzaku+Swift}, \citealt{2018A&A...616A..53L}).  The nature of the hard X-ray emission is not well-established, though it is thought to be attributable to the presence of an accretion disk: the hard X-rays have been variously modeled as isobaric cooling flows \citep{2007ApJ...671..741L}; or partially covered \citep{2009ApJ...701.1992K} or clumpy absorption of thermal emission \citep{2021MNRAS.500.4801D}; or post-shock regions above the polar caps of a magnetized white dwarf \citep{2016A&A...592A..58D}, and are able to explain the features of the spectrum at high energies ($E{\gtrsim}$3~keV).

\rtcru\ also exhibits several variability features like aperiodic flickering at timescales of a few~ks, and a strong correlation of spectral hardness with overall brightness, with higher intensities corresponding to softer spectra \citep[see][]{2021MNRAS.500.4801D}.  Such features are characteristic of emission driven by changes near the inner boundary of the accretion disk.  The question that arises then is what the origin of this variability could be.  As noted above, the observed X-ray spectrum has been variously modeled as an intrinsic change in the soft thermal emission component as well as changes in a continuum component due to intervening absorption.  The presence of spectral lines during increases of soft flux, especially if they are the dominant contributors to the soft emission, would support the former scenario, while the lack of such lines would favor the latter scenario.  The \chandra/LETGS+HRC-S data we focus on below (Section~\ref{s:data}) was originally obtained to settle this question, but the analysis was limited because of the relatively high background encountered.  While an adequate spectral fit that included partial covering of a thermal component was obtained by \citet{2021MNRAS.500.4801D}, the fit was driven by the higher energy signal.  Thus, the nature of the soft emission remained unsettled.

\subsection{Data collection: the \chandra/LETGS+HRC-S Observation}\label{s:data}

\begin{figure}
    \centering
    \includegraphics[width=\linewidth]{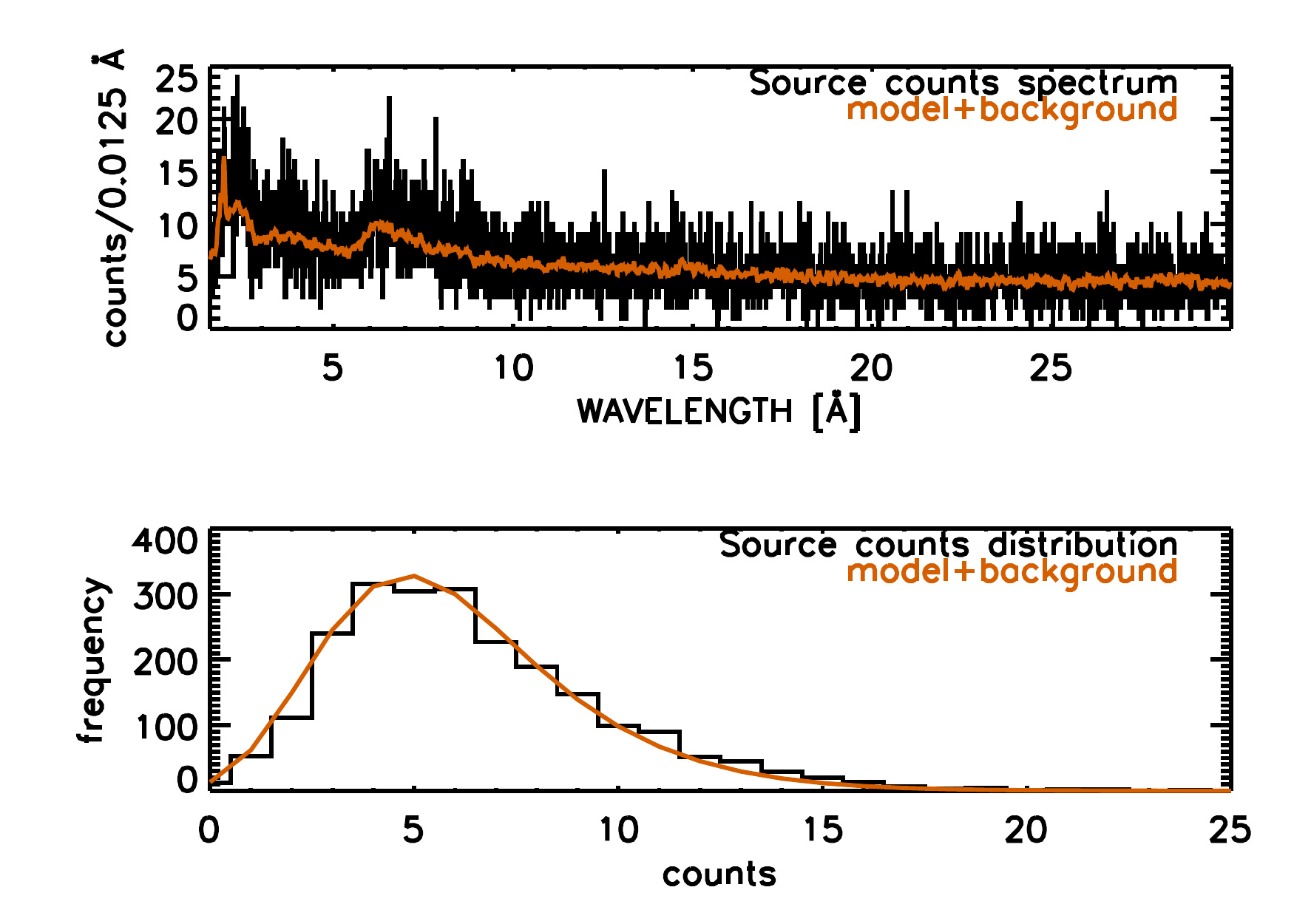}
    \caption{Comparing the spectrum of \rtcru\ with the model from \citet{2021MNRAS.500.4801D}.  {\sl Top:} The \chandra/LETGS+HRC-S counts spectrum is shown as the black histogram, and the sum of the estimated background and the predicted model are shown in red. {\sl Bottom:} The distribution of counts in each bin in the observed spectrum is shown as the black histogram, and the expected Poisson distribution based on the model intensity and the estimated background is shown as the red curve.
    }
    \label{fig:modelfits}
\end{figure}

\rtcru\ has been observed extensively with both {\sl Swift/XRT} and the \chandra\ gratings.  Here we focus specifically on the \chandra/LETGS+HRC-S observations that were carried out in Nov 2015 for a total of 78.9~ks (PI: M.\ Karovska; ObsIDs 16688 and 17810) and whose analysis was described by \citet{2021MNRAS.500.4801D}.
\citeauthor{2021MNRAS.500.4801D} modeled the spectrum as a combination of a power-law component (with index $\Gamma\approx$1.7), a thermal soft excess (with temperature $\approx$1.3~keV), and an ad hoc triplet of emission lines from Fe~K$\alpha$, Fe~XXV, and Fe~XXVI in the 6-7~keV region. 
The observed spectrum is shown in the top panel of Figure~\ref{fig:modelfits} in the $\lambda$=[1.56,30]~\AA\ wavelength range (corresponding to photon energies $E=[0.41,8]$~keV) as the black histogram, along with the best-fit astrophysical model spectrum (together with the estimated background contribution) as the red curve. The hard emission in the 6-7 keV range likely originates close to the accretor, e.g., from a region of the boundary layer of the accretion disk or a bright spot.  It could also be associated with collisional excitation in the vicinity of the WD accretor, including in the inner-jet regions (e.g., \citealt{2009ApJ...701.1992K}, \citealt{2010ApJ...710L.132K}, \citealt{2014MNRAS.437..857E}).

\begin{table*} 
\begin{tabular} 
{ 
	|c|cc|cc|
	c|c|
	}
		\hline
	Passbands & \multicolumn{2}{c|}{Passband range} & \multicolumn{2}{c|}{Dominant lines and Features} & $N_{r}$&  $n_{r}$\\
	of interest ($\passband{r}$)& [\AA] & [keV]& \hfil & [\AA]& &  \\
		\hline
	$\passband{1}$& 1.65-2.05 & 6.05-7.5 &	Fe~K & 1.785/1.853/1.944& 6879  & 368  \\
	$\passband{2}$&	5.5-10.2  & 1.2-2.25 & Ir-M edge & $\approx$6.5 &75699 & 3311 \\
	$\passband{3}$& 11.5-13.0 & 0.95-1.08 & Ne\,X &12.131 &22809      & 730 \\
	$\passband{4}$&	14.6-15.15 & 0.82-0.85 &Fe\,XVII &15.014& 8354   & 247   \\
    $\passband{5}$&	16.2-17.4 & 0.71-0.765 &Fe\,XVII  &17.051 & 17331    & 471  \\
	$\passband{6}$&	18.5-19.5 & 0.65-0.67 &O VIII  &18.967 &14060   & 390    \\
	$\passband{7}$&	21.3-21.75 & 0.57-0.582 &O VII(r) &   21.602& 6192 &  179   \\
	$\passband{8}$&	21.65-22.0 & 0.564-0.573 &O VII(i) & 21.804& 4927  & 145 \\
	$\passband{9}$&	21.9-23.0 & 0.539-0.566 & O VII(f) & 22.101&15255  & 390    \\
		\hline
	\end{tabular}
	\caption{Summary of wavelength bands of interest. The second and third columns of the table report the wavelength range (in both \AA \ and keV) of each of the regions, $\passband{r}$, $r=1,\dots,9$, considered. The expected spectral lines and their expected positions (in \AA) on each of the regions are listed in the fourth and fifth columns. The size of the source-free sample $(N_{r})$  and the physics sample $(n_{r})$ for each of the region $r$ are given in the last two columns.
    } 
    \label{table:bands}
\end{table*}

However, the soft component of this model is not well constrained, partly due to the lack of sensitivity of Swift and \chandra/ACIS detectors at $E{\lesssim}2$~keV and the high instrument background present in \chandra/LETGS+HRC-S.  The advantage of high spectral resolution observations is that emission lines, if they exist, can be located, and their presence and identification can provide useful diagnostic information about the emitting plasma.  The high background in the LETGS+HRC-S spectrum makes this a difficult problem.  The upper panel of Figure~\ref{fig:modelfits} shows several spikes that could be tagged as spectral lines, but the question arises as to how significant each of those identifications would be.  Methods that are typically used in these situations compare the heights of the spikes to the baseline intensity levels, and flag them as significant if they exceed a certain threshold.  But the large number of bins present in high-resolution spectra implies that the chances of false positives being flagged as a line are high.  This is demonstrated in the bottom panel of Figure~\ref{fig:modelfits}, which compares the observed distribution of counts in the bins (black stepped histogram) with the expected Poisson distribution derived from the model and background (smooth red curve).  Typically, a threshold is set based on the expected distribution such that a tolerable number of false detections are accepted; this tolerance can be set as accepting one false detection over the sample \citep[as is done in, e.g., {\tt wavdetect};][]{2002ApJS..138..185F}.  There are $\approx$2280 bins in the observed spectrum shown in the upper panel, yielding $p<0.000044$ as the required threshold value (note that the usual "$3\sigma$" detection criterion often used in astronomy corresponds to $p<0.0027$, and adopting it would result in as many as 6 false claims of detection; this issue will worsen when higher resolution spectra covering larger wavelength ranges are obtained), which translates to the requirement that a fluctuation exceeds 20 counts before it can be considered significant.

\begin{figure*}
    \centering
    \vspace{-1.5cm}
    \includegraphics[width=\linewidth]{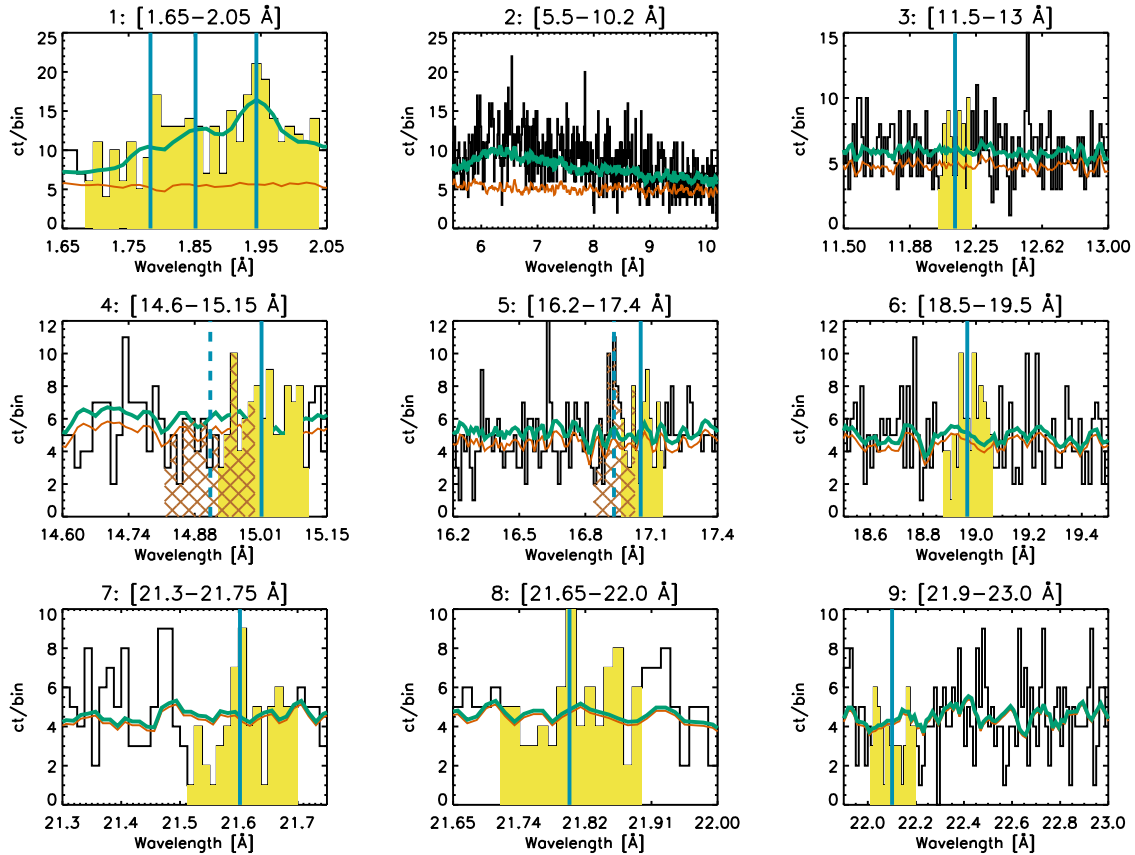}
        \vspace{-1.5cm}
    \caption{The counts spectrum of \rtcru\ (Figure~\ref{fig:modelfits}), without background subtraction, zoomed in to the various bands of interest (see Table~\ref{table:bands}).  The counts spectrum is shown as the black histogram, with the best-fit model spectrum from \citet{2021MNRAS.500.4801D} shown as the green curve.  The estimated smoothed background is shown as the red curve.  The blue vertical lines indicate the lines of interest in each band (except for $\passband{2}$, which covers an instrument feature).  The shaded yellow regions cover the approximate expected widths of the line response for each of the lines studied here.  The brown cross-hatched regions in passbands $\passband{4}$ and $\passband{5}$ mark serendipitous testing locations (see Discussion).}
    \label{fig:fullbands}
\end{figure*}

The goal of this observation was to obtain a spectrum at wavelengths $\lambda{\gtrsim}6$~\AA\ (energies $E{\lesssim}2$~keV), where the sensitivity of the instrument is higher on average than previous observations, in particular at the locations dominated by Ne~X~$\lambda$12.13 and the two Fe~XVII lines at $\lambda\lambda$15.01,17.05.  In addition, we also consider other wavelength ranges like the Fe~K region (where spectral lines have indeed been detected and modeled) and wavelength regions where O~VIII~$\lambda$18.96 and the O~VII He-like triplets at $\lambda\lambda$21.6,21.8,22.1 are to be found.  These wavelength bands of interest are denoted with $\passband{r}$, $r=1,\dots,9$ and are listed in Table~\ref{table:bands}; the Table shows the wavelength range, the dominant structure expected in these bands, and the number of events within the passbands in both the source region ($n_{r}$) as well as the source-free background region ($N_{r}$, collected in an area 39.4$\times$ greater than the source region).  The observed spectra (black histograms) and model predicted spectra (green curves) are shown, along with the expected background (red curves), zoomed in to these wavelength regions in Figure~\ref{fig:fullbands}; the expected locations of spectral lines of interest are marked with yellow shaded regions centered around vertical green lines.

Inspection of the spectral regions of interest in Figure~\ref{fig:fullbands} shows hints that some emission structure may be present in some cases, e.g., in passbands $\passband{3}$ (Ne\,X), $\passband{5}$ (Fe\,XVII), $\passband{6}$ (O\,VIII), $\passband{7}$ (O\,VII(r)), and $\passband{8}$ (O\,VII(i)).  
We carry out a naive estimation of the line intensities by counting the number of photons within $\pm$0.1\AA\ (corresponding to the width of the line response) at the putative locations of the lines, collected within both the source and background datasets, and computing the Bayesian posterior density of the line intensities \citep[viz.][]{2014ApJ...796...24P}.  Table~\ref{tab:modelct} lists the mode and the 68\% highest posterior density (HPD) intervals for the resulting posterior distributions for each line or band of interest.  Several line intensities are estimated formally at significance $>1\sigma$.  An analysis that ignores the problem of multiple tests would claim detections of these lines, but it is possible that such claims would turn out to be false positives.  A more careful and principled method must be used to evaluate their reality.
We, therefore, employ the method proposed in \citep[][]{algeri2020,algeri2021}  to adequately model background shapes on each of the regions considered. 
\begin{table}
    \centering
        \begin{tabular}{c c c}
    \hline\hline
    Wavelength & Model$^a$ & Estimate$^b$ \\
    Band & Expected & Observed \\
    \hline
    $\passband{1}$ & 175.7 & 186 $^{<205}_{>167}$ \\
    $\passband{2}$ & 1021.0 & 1370 $^{<1430}_{>1310}$ \\
    $12.131{\pm}0.1$~\AA~$\subset\passband{3}$ & 16.4 & 20.0 $^{<30.6}_{>10.9}$ \\
    $15.014{\pm}0.1$~\AA~$\subset\passband{4}$ & 12.7 & 20.6 $^{<31.2}_{>11.4}$ \\
    $17.051{\pm}0.1$~\AA~$\subset\passband{5}$ & 9.15 & 5.6 $^{<15.2}_{>1.4}$ \\
    $18.967{\pm}0.1$~\AA~$\subset\passband{6}$ & 6.78 & 16.1 $^{<26.3}_{>7.7}$ \\
    $21.602{\pm}0.1$~\AA~$\subset\passband{7}$ & 2.8 & 0 $^{<6.7}_{\ge0}$ \\
    $21.804{\pm}0.1$~\AA~$\subset\passband{8}$ & 2.6 & 9.8 $^{<19.5}_{>3.5}$ \\
    $22.101{\pm}0.1$~\AA~$\subset\passband{9}$ & 2.5 & 0 $^{<4.8}_{\ge0}$ \\
    \hline
    \multicolumn{3}{l}{$a$: Expected model counts based on \citet{2021MNRAS.500.4801D}} \\
    \multicolumn{3}{l}{$b$: Mode of the posterior density distribution of source} \\
    \multicolumn{3}{l}{~~~~ intensity \citep[][]{2014ApJ...796...24P} and the 68\% HPD } \\
    \multicolumn{3}{l}{~~~~ (highest posterior density) uncertainty bounds} \\
    \end{tabular}
    \caption{Model predicted and estimated counts at line locations of interest.}
    \label{tab:modelct}
\end{table}
Moreover, since our analysis requires us to test for the presence of one or more lines on each of  the above-mentioned regions, adequate corrections for multiple hypotheses testing are also implemented. Such step is necessary to ensure that the overall probability of a false discovery across any of the regions considered does not exceed the desired significance level. 
Finally, we also test whether the feature present at 16.93~\AA\ in $\passband{5}$, shortward of the Fe~XVII $\lambda$17.04 line by $\approx$2100~km~s$^{-1}$ is real, in the sense of being detectable.

We find that none of the hypothesised thermal spectral lines listed in Table~\ref{fig:fullbands} are detectable (see Section~\ref{sec:signaldetection}), and thus we place upper limits on the strengths of the lines that are well above reasonable variations in the fitted spectral model (cf.\ expected model counts listed in Table~\ref{tab:modelct}).  In contrast, we find that the serendipitously identified line at 16.93~\AA\ is detectable with a corrected $p$-value $\approx$0.01.  We discuss the possible origins of this line in Section~\ref{sec:discuss}.

\section{Statistical Methods}\label{s:methods}
As outlined in Sections \ref{s:intro}-\ref{sec:rtcru}, the presence of high background makes it necessary to adequately characterise its shape and discern its characteristic features from random fluctuations. We show here that this can be done by means of \emph{smooth models} and \emph{smooth tests} originally introduced in statistical literature by \citet{neyman1937}. 

In the context of the analysis of high-resolution spectra, smooth tests are particularly useful in that they naturally integrate estimation and inference to adequately handle uncertain background shapes. Moreover, when aiming to detect new signals, they provide a trade-off between  tests of hypotheses and goodness-of-fit tests.  Specifically, while tests of hypotheses, such as the Likelihood Ratio Test (LRT) \citep[e.g.,][]{wilks}, are known to have high power towards a well-specified alternative model, their validity is severely undermined if the models under comparison are misspecified. In our case, this would occur, for instance, when searching for a few strong spectral line at the wrong position, or when the deviations from the background are due to the signal of a continuous, widely spread source. A classical approach to obviate this problem is that of relying on classical goodness-of-fit tests, such as Kolmogorov-Smirnov \citep{kolmogorov,smirnov},  or Pearson's $\chi^2$ \citep{pearson}, and which are expected to have some power against all possible deviations from the background model. However, since different tests have higher power towards different alternatives,  they may not be sensible in detecting deviations towards the desired direction. Smooth tests, on the other hand, enjoy high power only towards a finite number of possible directions. Interestingly, the user can rely on suitable data-driven procedures to select  the most ``significant'' directions towards which deviations from the background  occur. Therefore, they effectively let the data decide ``where to look''. 

From a methodological perspective, the scope of our analysis is three folds. Our first goal is to assess the validity of our postulated background model and, when needed, provide a data-driven correction for it. Second,  we aim to determine whether the Chandra LETGS+HRC-S Data provides significant evidence to conclude that spectral lines are present in addition to the background. Third, in regions where no significant deviations from the background model are detected, we proceed by setting upper limits on the intensity of the expected signals.
In order to achieve these goals, we rely on suitably constructed \emph{smooth tests} and likelihood-ratio tests. 

\subsection{Smooth Models and Smooth Tests} \label{sec:3.1}
Smooth tests are an inferential procedure for model assessment originally introduced by \cite{neyman1937}. In the context of 
astronomical spectra,
they are particularly advantageous because, in addition to validating the postulated astrophysical models, they also provide valuable insights on how misspecified models can be improved by means of smooth models. The latter consist of an ``updated'' version of the initial proposed model
on the basis of the data observed. Sections \ref{sec:3.1}-\ref{sec:adjust} outline the main steps characterizing these procedures. 

\subsubsection{Data-driven Model Calibration via Smooth Models}
\label{sec:3.1.1}

\begin{table}
\caption{Summary of the notation used throughout the manuscript.}
\begin{threeparttable}

\begin{tabular}{ll}
\hline
\hline
Symbol & Definition \\
\hline
Section \ref{sec:3.1}: \\
$X$ & Random variable with unknown distribution $F$\\
$x$ & The value of $X$ observed on the data \\
$F$ & Unknown cumulative distribution function (cdf) of $X$  \\
$f$ & Unknown probability density function (pdf) of $X$  \\
$G$ & Hypothesised cdf of $X$\\
$g$ & Hypothesised pdf of $X$\\
$d(u; G,F)$ &  ratio between the densities $f$ and $g$ evaluated at $u=G(x)$ \\
$\{h_j(u)\}_{j\geq 0}$ & A series of orthonormal basis
functions on $[0,1]$\\
$\theta_j$ & $j^{th}$ coefficient of the orthonormal expansion for $d(u; G,F)$\\
$\widehat{\theta}_j$ & Estimate of $\theta_j$\\
$m$& Number of terms used in the expansion for $d(u; G,F)$\\
$\widehat{d}(u;G,F)$ & Estimate of $d(u;G,F)$\\
$\widehat{f}$& Smooth estimator of  $f$\\
$D_m$ & Deviance statistic\\
$d_m$ & The value of $D_m$ observed on the data \\ 
$K_m$& K-statistic \\
$k_m$ & The value of $K_m$ observed on the data \\ 
$\alpha$ & A pre-specified significance level \\
Section \ref{sec:3.2}: \\
$M$& Upper bound for the selection of $m$\\
$\widehat{\theta}_{(j)}$ &  The $jth$ largest (in magnitude) among the coefficients $\theta_j$\\ 
Section \ref{sec:adjust}: \\
$D_{(m)}$ & Post-selection Deviance statistic\\
$d_{(m)}$ & The value of $D_{(m)}$ observed on the data \\ 
$K_{(m)}$ &Post-selection K-statistic \\
$k_{(m)}$ & The value of $K_{(m)}$ observed on the data \\ 
Section \ref{sec:background_only}:\\
$\passband{r}$& $r$th wavelength band of interest, $r=1,...,9$\\ 
$B_{\passband{r}}$ &True unknown background distribution for region $\passband{r}$ \\
$b_{\passband{r}}$ &True unknown background density for region $\passband{r}$ \\
$g_{\passband{r}}$& Postulated uniform background density for region $\passband{r}$ \\
$\widehat{B}_{\passband{r}}$ & Corrected background distribution for region $\passband{r}$\\ 
$\widehat{b}_{\passband{r}}$ & Corrected background density for region $\passband{r}$\\ 
$C_w$ & $w$th calibration region, $w=1,...,5$\\ 
$\widehat{b}_{C_w}$ & Corrected background density for combined region $C_w$\\ 
$N_{r}$ & The size of the source-free sample for the region $r$\\ 
$n_{r}$ & The size of the physics sample for the region $r$\\ 

Section \ref{sec:4.2}: \\
$F_{r}$ &Unknown distribution for physics sample for region $\passband{r}$ \\
$f_{r}$ &Unknown density for physics sample for region $\passband{r}$ \\
Section \ref{sec:constructUL}: \\
$s_{r}$ &Density of spectral line for region $\passband{r}$, $r=3,...,9$ \\
$\eta_r$ &Relative intensity of the expected signal for region $\passband{r}$\\
$q_{r}$ &Density of physics sample for region $\passband{r}$ \\
\hline
\end{tabular}
\end{threeparttable}
\vspace{0.5mm}
\end{table}

Let $X$ be a continuous random variable with an unknown cumulative distribution function (CDF) $F$, and an unknown probability density function (PDF) $f$. $F$ corresponds to the true distribution from which our data were generated.  Despite $F$ being unknown, we aim to assess if it can be reasonably approximated by a hypothesised distribution $G$, with PDF $g$. The goal is to test the hypotheses
\begin{equation}
H_0: F = G; \quad vs \quad H_1: F \neq G
    \label{eqn:h00}
\end{equation}
To perform the test in (\ref{eqn:h00}), we begin by rewriting the true density, $f$, as a function of the postulated model $g$, i.e.,
\begin{equation}
\begin{aligned}
    f(x)=g(x)d\bigl(G(x);G,F\bigl)
    \label{eqn:skewG}
\end{aligned}
\end{equation}
where $d\bigl(G(x);G,F\bigl)=\frac{f(x)}{g(x)}$; that is,  $d\bigl(G(x);G,F\bigl)$ corresponds to the density ratio between  $f$ and $g$ and can be expressed in the quantile domain as  
\begin{equation*}
    d(u;G,F) = \frac{f\bigl(G^{-1}(u)\bigl)}{g\bigl(G^{-1}(u)\bigl)} \quad \text{with } u = G(x)\in[0,1]
\end{equation*}
where $G^{-1}(u)$ is the quantile function of $X$ under model $G$. In statistics literature, $d(u;G,F)$ is also referred to as "comparison density" \cite[e.g.][]{parzen2004}; such nomenclature is used to emphasise that $d(u;G,F)$ is the PDF of the random variable $U=G(X)$ \cite[see proposition 3.1 in][]{algeri2021}. 

Under the assumption that the density  $d(u;G,F)$ is square integrable on the unit interval, we can represent it via a series of orthonormal basis functions $\{ h_j(u)\}_{j\geq 0} $, i.e., 
\begin{equation}
d(u;G,F) = 1+ \sum_{j=1}^{\infty} \theta_j h_j(u) 
\label{eqn:3}
\end{equation}
with $h_0(u)=1$ and  $\theta_j =\int_0^1 h_j(u)d(u;G,F) du$. 
The choice of the basis functions $h_j$ in (\ref{eqn:3}) is arbitrary. For example,  in virtue of their simple implementation and generalisability to the discrete setting, a popular choice are the normalised shifted Legendre polynomials \citep[e.g.,][]{neyman1937, Led1994}. 

A \emph{smooth model} can then be specified by truncating the series in \eqref{eqn:3} at a point $m$. Moreover, once a set of independent and identically distributed observations, $x_1,..., x_n$, has been collected, the coefficients $\theta_j$ are estimated via
\begin{equation}
\widehat{\theta}_j = \frac{1}{n}\sum_{i=1}^{n} h_j\bigl(G(x_i)\bigl) = \frac{1}{n}\sum_{i=1}^{n} h_j(u_i).
    \label{eqn:coef}
\end{equation}
It follows that an estimator of $d(G(x);G,F)$ in (\ref{eqn:3}) is
\begin{equation}
\widehat{d}\bigl(G(x);G,F\bigl) = 1+\sum_{j=1}^{m} \widehat{\theta}_j h_j\bigl(G(x)\bigl).
    \label{eqn:den_estimate}
\end{equation}
Finally, a \emph{smooth estimator} of the true density $f$ is
\begin{equation}
\widehat{f}(x) = g(x)\widehat{d}\bigl(G(x);G,F\bigl) = g(x)\Bigl[1+\sum_{j=1}^{m} \widehat{\theta_j} h_j\bigl(G(x)\bigl)\Bigl].
    \label{eqn:5}
\end{equation}

As described in Section \ref{sec:3.1.2}, this modeling strategy naturally leads to the inferential framework needed to test the hypotheses in \eqref{eqn:h00}.  

\subsubsection{Smooth Tests} \label{sec:3.1.2}
We begin by considering the simplified scenario where the point of truncation, $m$, in \eqref{eqn:den_estimate} is fixed and all the first $m$ coefficients are included in our estimators in \eqref{eqn:den_estimate} and \eqref{eqn:5}. This simplifying assumption allows us to introduce the main asymptotic results at the core of our inferential strategy. A detailed discussion on model selection and adequate inferential adjustments 
is postponed to Sections \ref{sec:3.2}-\ref{sec:adjust}.

The decomposition in ($\ref{eqn:skewG}$) allows us to rewrite the hypotheses in ($\ref{eqn:h00}$) as
\begin{equation}
\begin{split}
H_0: d\bigl(u;G,F\bigl) &= 1\quad \text{for all $u\in[0,1]$}, \quad \text{versus}\\ \quad H_1:d\bigl(u;G,F\bigl) &\neq 1\quad \text{for some $u\in[0,1]$}
    \label{eqn:h0}
    \end{split}
\end{equation}
Furthermore, by exploiting our estimator in $(\ref{eqn:den_estimate})$, we can reformulate \eqref{eqn:h0} as
\begin{equation}
    \begin{aligned}
    &H_0: \theta_1=...=\theta_m=0 \quad \text{versus} \\
    &H_1: \theta_j \neq 0  \text{ for at least one } j=1,...,m
\label{eqn:h11}
    \end{aligned}
\end{equation}
Notice that $H_0$ in (\ref{eqn:h0}) implies $H_0$ in (\ref{eqn:h11}) but $H_1$ in (\ref{eqn:h0}) does not imply $H_1$ in (\ref{eqn:h11}). Whereas, $H_1$ in (\ref{eqn:h11}) implies $H_1$ in  (\ref{eqn:h0}) and thus by testing $H_0$ and $H_1$ in (\ref{eqn:h11}) we can determine if the true distribution $F$ deviates significantly from our hypothesised distribution $G$. 

To test the hypotheses in (\ref{eqn:h11}), we rely on two different test statistics. A common choice in the context of smooth tests \citep[e.g.][]{Led1994,Led1997} is the deviance statistic, i.e.,
\begin{equation}
D_m=n\sum_{j=1}^m \widehat{\theta_j}^2;
\label{eqn:devstat}
\end{equation}
whereas, in this manuscript, we propose, in addition to \eqref{eqn:devstat},  the use of the \emph{K-statistic} which specifies as
\begin{equation*}
K_m = \mathop{\max}_{j=1,..,m} n\widehat{\theta_j}^2.
\end{equation*}
It is easy to show \cite[e.g.][]{algeri2020} that, under $H_0$, and as $n\to\infty$, the estimators $\widehat{\theta_j}$ in \eqref{eqn:coef} converge in distribution to normally distributed random variable with mean zero and variance $1/n$.  It follows that
\begin{equation*}
D_m= \sum_{j=1}^m (\sqrt{n}\widehat{\theta}_j)^2 \stackrel{d}{\rightarrow} \chi^2_m.
\end{equation*}
Hence, an asymptotic p-value for the deviance test is
\begin{equation}
     \text{p-value}= P(\chi^2_m >d_m) \quad 
    \label{eqn:pval}
\end{equation}
where $d_m$ is the value of $D_m$ observed on the data. 

\vspace{3mm}

The asymptotic distribution of the K-statistic under $H_0$ can be  derived by noticing that, asymptotically, $K_m$ is distributed as the maximum of $m$  $\chi^2_1$ distributed random variables. Therefore, letting $k_m$ be the observed value of $K_m$ on the data, its asymptotic p-value is
\begin{equation}
\begin{aligned}
\text{p-value} = 1-P(\chi^2_1 \leq k_m)^{m}.
    \label{eqn:kpval}
\end{aligned}
\end{equation}
The respective proof can be found in Appendix \ref{appendixA}. 

The null hypothesis is rejected when the p-value is smaller than a pre-specified significance level $\alpha$. For example, a discovery claim  at $2\sigma$ significance corresponds to a level $\alpha$ of approximately $0.05$. A rejection of the null hypothesis implies the postulated model $g$ deviates significantly from the true model $f$.

\subsection{Model Selection}\label{sec:3.2}
An important step of our analysis is that of selecting the basis functions to be included in our estimators (\ref{eqn:den_estimate}) and (\ref{eqn:5}). Here, we  rely on a model selection process based on the Bayesian information criterion (BIC) for selection \citep[e.g.,][]{deep2017,algeri2021}, and  can be summarised as follows:
\begin{itemize}
    \item[i.] Choose a suitably large value $M$ (usually 10).
    \item[ii.] Obtain the estimates $\widehat{\theta}_1,...,\widehat{\theta}_{M}$ as in (\ref{eqn:coef}).
    \item[iii.] Rearrange $\widehat{\theta}_j^2$ in decreasing order, i.e.,
    \begin{equation*}
    \begin{aligned}
    \widehat{\theta}^2_{(1)} \geq \widehat{\theta}^2_{(2)} \geq ... \geq \widehat{\theta}^2_{(M)}, \\
    \text{ with } \widehat{\theta}^2_{(j)} \text{ denoting the } jth \text{ largest squared coefficient. }
    \end{aligned}
    \end{equation*}
    \item[iv.] Choose the largest $m$ that maximises 
    \begin{equation}
    \begin{aligned}
    BIC(m) = \sum_{j=1}^m \widehat{\theta}^2_{(j)} - \frac{m\log n}{n}.
    \end{aligned}
    \label{eqn:BIC}
    \end{equation}
\end{itemize}
The procedure outlined here is data-dependent, that is, the value of $m$ selected may vary over different samples. It is therefore necessary to account for the randomness associated with the model selection process. This can be done as described in Section \ref{sec:adjust}.

\subsection{Post-selection Inference Adjustments}\label{sec:adjust}

Traditional inference is typically constructed under the assumption that the model under study has been selected independently from the data available. In many practical scenarios, however, a data-driven  selection procedure is typically implemented, and thus,  classical inferential results fail to hold due to the randomness associated with the selection process. While one can easily overcome this problem by relying on data splitting \cite[e.g.][]{moran1973, cox1975}, or bootstrapping  \cite[e.g.][]{sara2021}, in our setting, it is possible to identify suitable post-selection adjustments for the p-values in \eqref{eqn:pval} and \eqref{eqn:kpval}. 

For what concerns the deviance statistic, a post-selection adjusted p-value can be constructed as described in \citet{algeri2021}. Specifically,  let 
\begin{equation*}
D_{(m)} = \sum_{j=1}^m \widehat{\theta}_{(j)}^2
\end{equation*}
be the deviance statistic obtained as the sum of the squares of the $m$ largest estimated coefficients, with $m$ selected via the BIC criterion in \eqref{eqn:BIC}, and denote with $d_{(m)}$ be its value observed on the data.  An adjusted  p-value for the test in \eqref{eqn:h11} is 
\begin{equation}
\text{p-value}_{adj,M} = P\bigl(\chi^2_M > d_{(m)}\bigl).
\label{eqn:adjpvalnaive}
\end{equation}
Notice that, conversely from (\ref{eqn:pval}), in (\ref{eqn:adjpvalnaive}) the observed value of the deviance statistic is compared to a $\chi_M^2$ rather than $\chi_m^2$. In virtue of its conservatives, we refer to the correction in (\ref{eqn:adjpvalnaive}) as "naive correction". 

An alternative approach consists of applying the usual Bonferroni’s correction \cite[e.g.][]{m1977} and typically used in the context of multiple hypothesis testing. In this case, the  adjusted deviance p-value is given by
\begin{equation}
\text{p-value}_{adj, B}= \max\Bigl\{M \cdot P\bigl(\chi^2_m >d_{(m)}\bigl),1\Bigl\}
\label{eqn:bonadjpval}
\end{equation}

Finally, the post-selection  K-statistic specifies as 
\begin{equation*}
K_{(m)} = \max_{j=1,...,m} n\widehat{\theta}_{(j)}^2,
\end{equation*}
where $m$ is the value which maximises the BIC in (\ref{eqn:BIC}), and let $k_{(m)}$ be the value of $K_{(m)}$ observed on the data. 
A post-selection adjustment for the respective p-value can be implemented as formalised in Theorem \ref{thm:1}. The respective proof is given in the Appendix \ref{appendixB}.

\newtheorem{theorem}{Theorem}
\newtheorem{definition}{Definition}
\begin{theorem} \label{thm:1}
As $n \to \infty$, an upper bound for the limit of $P(K_{(m)}\geq k_{(m)}|H_0)$  is
\begin{equation}
\text{p-value}_{adj,K} =1- P\bigl( \chi^2_1\leq k_{(m)}\bigl)^M.
\label{eqn:kadjpval}
\end{equation}
\end{theorem}
A comparison in terms of power and type I error of tests based on \eqref{eqn:adjpvalnaive}, \eqref{eqn:bonadjpval} and \eqref{eqn:kadjpval} is postponed to Section \ref{sec:constructUL}.

\section{Statistical Data Analysis of \rtcru}\label{s:analysis}

Here we describe how smooth tests can be implemented to thoroughly study the $X$-ray spectra of \rtcru. As outlined in Section \ref{s:intro}, our strategy is that of conducting a separate analysis on each of the 9 wavelength bands listed in Table~\ref{table:bands}.  

Specifically, in Section~\ref{sec:background_only}, we first demonstrate   how the shape of the background can be  extracted  on the basis of the information contained in the source-free, background-only dataset.  Second, in Section~\ref{sec:signaldetection}, we  apply the same method to the source spectrum. This step allows us to determine whether a significant difference can be established between the source spectrum and the background spectrum, without imposing any distributional assumption on potential signals.  Third, if no difference can be detected, in  Section~\ref{sec:LRTUL} we complement our analysis by means of Likelihood Ratio Tests (LRTs) to assess for the presence of the spectral lines listed in Table \ref{table:bands}. For both smooth tests and the LRTs, adequate adjustments for multiple hypothesis testing are  implemented in order to control for the probability of a false discovery across all the regions of the spectrum being tested. The statistical properties of the testing procedures considered are investigated in Section~\ref{sec:constructUL}. Finally, since no spectral line is detected,  we set upper limit on the counts necessary for a detection in each of the regions considered.

\subsection{Data-driven Background Corrections} \label{sec:background_only}
When assessing the validity of a postulated background model and implementing adequate data-driven corrections, it is necessary to ensure the size of the source-free sample considered is sufficiently large to reduce the  uncertainty associated with the estimation of the background model. Therefore, here we proceed to combine the regions of interest $\passband{r}$ in Table \ref{table:bands} into five ``calibration'' intervals $C_w$, $w=1,\dots,5$, and defined as in Table \ref{table:var-importance2}.

The choice of considering larger regions ensures that, for each of the newly constructed calibration intervals $C_3$-$C_5$ considered, the corresponding source-free sub-sample includes at least 20,000 events. 

The background is assumed to be flat over the entire search area and thus we proceed by testing this assumption on each sub-region by means of the tools described in Section \ref{s:methods}.
The adjusted p-values to test  the uniformity of the background, for each calibration region $C_w$, $w=1,\dots,5$ have been computed as in (\ref{eqn:adjpvalnaive}), (\ref{eqn:bonadjpval}) and (\ref{eqn:kadjpval}). The results are summarised in Table $\ref{table:var-importance2}$. 

In three out of the five calibration regions in Table \ref{table:var-importance2}, the data distribution   is consistent with the uniform model (the adjusted p-values are all equal to one). Whereas on $C_2$ and $C_4$, the flat background model is rejected. Therefore, we proceed to implement a data-driven correction for them as described in Section \ref{sec:3.2}. The newly estimated background densities are 
\begin{equation}
    \begin{aligned}
&\widehat{b}_{C_2}(x)=  0.2424 - 0.0038x,\quad x\in [5.5,10.2] \ \text{\AA} , \\
&\widehat{b}_{C_4}(x)=  1.3994 - 0.0250x,\quad x\in [14.6,17.4] \ \text{\AA} . 
    \label{eqn:corre}
\end{aligned}
\end{equation}

Our ultimate goal is to detect signals over the regions of interest, $\passband{r}$, defined as in Table \ref{table:bands},  it is, therefore, necessary to convert the re-calibrated background densities obtained for $C_2$ and $C_4$ into corrections for the background distribution over $\passband{r}$, $r=1,\dots, 9$. 

Specifically, let $B_{\passband{r}}$ be the true (unknown) background distribution on region $\passband{r}$ and let $b_{\passband{r}}$ be its density. We denote with $\widehat{b}_{\passband{r}}$ our estimate of $b_{\passband{r}}$.
Since no deviation from uniformity was observed over $C_1,C_3$ and $C_5$, no background update was performed on the regions $\passband{1},\passband{3},\passband{6},\passband{7},\passband{8}$ and $\passband{9}$. Hence we set,
\begin{equation*}
\widehat{b}_{\passband{r}}(x) = \frac{1}{u_r-l_r}, \text{ for  $r=1,3,6,7,8,9$},
\end{equation*}
with $l_r$ and $u_r$ be, respectively, the lower and upper bounds of the wavelength range of region $\passband{r}$.
Whereas, we exploit \eqref{eqn:corre} to derive the newly calibrated background models for regions $\passband{2},\passband{4}$ and $\passband{5}$. 
Since $\passband{2}$ and $C_2$ coincide, we set $\widehat{b}_{\passband{2}}(x)=\widehat{b}_{C_2}(x)$ and defined  in \eqref{eqn:corre}.
Whereas, the background models for regions $\passband{4}$ and $\passband{5}$ are 
\begin{equation}
    \begin{aligned}
&\widehat{b}_{\passband{4}}(x)=  2.4749 - 0.0441x,\quad x\in [14.6,15.15] \ \text{\AA} , \\
&\widehat{b}_{\passband{5}}(x)=  1.1900 - 0.0212x,\quad x\in [16.2,17.4] \ \text{\AA} ,
    \label{eqn:corrected}
\end{aligned}
\end{equation}
and they have been derived as described in Appendix \ref{appendixD}. 
\begin{table*}
    \centering
	\begin{tabular}{
	|c|
	c!{\vrule width 1.4pt}
	c!{\vrule width 1.4pt}c|
	c|c!{\vrule width 1.4pt}c!{\vrule width 1.4pt}}
		\hline
		Combined  & Wavelength&m& Bonferroni & K & Naive & $N_{w}$ \\
		regions ($C_w$) &range in \AA & & (Sidak) & (Sidak) & (Sidak)& \\
		\hline
		$C_1$ & 1.65-2.05& 0 & 1.0000 (1.0000)& 1.0000 (1.0000)& 1.0000 (1.0000)&6879 \\
		$C_2$ & 5.5-10.2 &1 & 3.7687e-10 (1.8843e-09)&  3.7687e-10 (1.8843e-09)& 3.6797e-06 (1.8398e-05)&75699 \\
	    $C_3$ & 11.5-13.0& 0 & 1.0000 (1.0000)& 1.0000 (1.0000)& 1.0000 (1.0000)&22809 \\
		$C_4$ &  14.6-17.4&1 & 0.0004 (0.0020) & 0.0004 (0.0020)&  0.0796 (0.3396) & 41186 \\
		$C_5$ &  18.5-23.0&0 & 1.0000 (1.0000)& 1.0000 (1.0000)& 1.0000 (1.0000)& 63372 \\
		\hline
	\end{tabular}
		\caption{Summary of our background correction results for each calibration region. The first two columns are the combined calibration regions and corresponding combined wavelength ranges.	The third column reports the number, $m$, of coefficients selected out of $M=10$, for each of the combined regions considered. The  p-values adjusted for post-selection using the Bonferroni, K, and naive corrections and corrected for multiple hypothesis testing via Sidak  (see \ref{eqn:sidak}) are given in the fourth, fifth, and sixth columns, respectively. The size of the source-free sample $N_{w}$ for each of the combined regions $w=1,...,5$ are given in the last column.}
	\label{table:var-importance2}
\end{table*}

\begin{table*}
    \centering
	\begin{tabular}{
	|c!{\vrule width 1.4pt}
	c!{\vrule width 1.4pt}
	c|
	c|c!{\vrule width 1.4pt}}
		\hline
		Regions &m& Bonferroni & K & Naive \\
		of interest ($\passband{r}$) & & (Sidak) & (Sidak) & (Sidak) \\
		\hline
	$\passband{1}$ &  3 &0.0001 (0.0011)& 0.0071 (0.0397) & 0.0045 (0.0621)\\
	$\passband{2}$ &  3 & 1.0816e-18 (1.0817e-17)
	& 2.7907e-15 (2.9976e-14) & 3.3306e-15  (2.4980e-14)\\
	  $\passband{3}$ &  0& 1.0000 (1.0000)& 1.0000 (1.0000)& 1.0000 (1.0000)\\
	$\passband{4}$ & 0 & 1.0000 (1.0000)& 1.0000 (1.0000)&  1.0000 (1.0000)\\
	$\passband{5}$ & 0& 1.0000 (1.0000)& 1.0000 (1.0000)& 1.0000 (1.0000)\\
	$\passband{6}$ & 0& 1.0000 (1.0000)& 1.0000 (1.0000)& 1.0000 (1.0000) \\
	$\passband{7}$& 0 & 1.0000 (1.0000)& 1.0000 (1.0000)& 1.0000 (1.0000) \\
	$\passband{8}$ & 0& 1.0000 (1.0000)& 1.0000 (1.0000)& 1.0000 (1.0000)\\
	$\passband{9}$ & 0& 1.0000 (1.0000)& 1.0000 (1.0000)& 1.0000 (1.0000)\\
		\hline
	\end{tabular}
		\caption{
		Summary of our signal detection results  for each region of interest using smooth tests. The second column shows the number, $m$, of coefficients selected out of $M=10$, for each of the nine regions considered. The  p-values adjusted for post-selection using the Bonferroni, K, and naive corrections and corrected for multiple hypothesis testing via Sidak  (see \ref{eqn:sidak}) are given in the third, fourth and fifth columns, respectively. }
	\label{table:var-importance3}
\end{table*}
\begin{figure}
\includegraphics[scale=0.25]{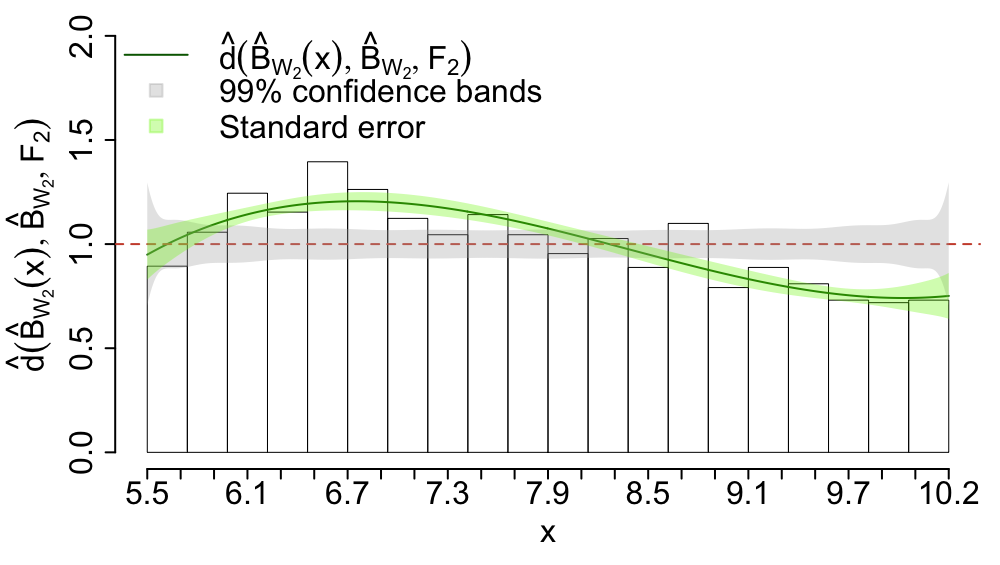}
\caption{Comparison density plot and histogram of the transformed physics data, $u_i=\widehat{B}_{\passband{2}}(x_i)$, $i=1,\dots,3311$, for region $\passband{2}$. The hypothesised model is the corrected background density $\widehat{b}_{\passband{2}}(x)$ in \eqref{eqn:corrected}. The estimated comparison density in (\ref{eqn:d3}) is displayed using a green solid line. The green bands are the estimated standard errors of $ \widehat{d}\bigl(\widehat{B}_{\passband{2}}(x);\widehat{B}_{\passband{2}},F_2\bigl)$ obtained by simulating samples from the estimator in (\ref{eqn:d3}). The red dashed line corresponds to the null hypothesis that $d\bigl(\widehat{B}_{\passband{2}}(x);\widehat{B}_{\passband{2}},F_2\bigl)$=1, i.e., the data distribution on $\passband{2}$ is not significantly different from the estimated background model $\widehat{b}_{\passband{2}}=\widehat{b}_{C_{2}}$ in \eqref{eqn:corre}. The grey bands correspond to the $99\%$ confidence bands under the null. The plot suggests that there exists a significant departure of the true data distribution from $\widehat{B}_{\passband{2}}(x)$, within the range of $x\in [6.1, 7.9]$\AA. }
\label{fig:region2}
\end{figure}
\begin{table}
    \centering
	\begin{tabular}{
	|c!{\vrule width 1.4pt}
	c|c|c|c!{\vrule width 1.4pt}}
		\hline
		Regions & Local &Sidak's  \\
			of interest ($\passband{r}$)& p-values&correction\\
		\hline
	    $\passband{3}$& 0.4810 &0.9899 \\
		$\passband{4}$& 0.1143 &0.5724 \\
		$\passband{5}$ & 0.3247 &0.9359\\
		$\passband{6}$& 0.0385 &0.2402\\
	$\passband{7}$ & 0.2612 &0.8799\\
	$\passband{8}$  & 0.5000 & 0.9922\\
	$\passband{9}$ & 0.5000 &0.9922\\
		\hline
	\end{tabular}
      \caption{Local P-values  and adequate multiple hypothesis testing adjustments when testing for spectral lines via LRTs.}
	\label{table:likelihood}
\end{table}

\subsection{Signal Detection via Smooth Tests and LRTs} \label{sec:4.2}

\subsubsection{Testing multiple regions and adequate corrections}
\label{sec:MHT}
Let $F_r$ be the true distribution from which the observations in the source sample were generated over region $\passband{r}$, $r=1,...,9$, and denote with $f_r$ its density. We assume that, if no signal is present, the true density $f_r$ is well approximated by our background estimate $\widehat{b}_{\passband{r}}$ obtained as in Section \ref{sec:background_only}. Hence, in Section \ref{sec:signaldetection}, we employ smooth tests to assess  if significant deviations from $\widehat{b}_{\passband{r}}$ occur on any of the nine regions considered.  For regions where no significant departure from the background model is detected, we will proceed with a thorough search for spectral lines via  LRTs, as described in Section \ref{sec:LRTUL}.

Notice that, in both our smooth tests and LRTs analyses, multiple tests are conducted simultaneously over different regions. It is therefore
necessary to correct the resulting p-values for multiple hypotheses testing in order to ensure that the probability of a false discovery across the entire spectrum is no larger than the predetermined $\alpha$. Hereinafter, we will refer to the latter as the \emph{global significance} level. 

Under the hypothesis of independence of the tests being conducted, $\alpha$ relates to the \emph{local significance} level, $\alpha_r$, of incorrectly rejecting the background-only hypothesis on region $W_r$ in that
\begin{align*}
    \alpha=1-\prod_{r=1}^R(1-\alpha_r).
\end{align*}
The reader is directed to \citet[][]{algeri2016} for a self-contained review on the problem of multiple hypothesis testing in searches for new physics also referred to in physics literature as \emph{look-elsewhere effect}.

In our setup, since the regions in which our spectrum is divided are non-overlapping, we can assume independence among the tests performed separately on each of them. Hence, we can adequately adjust the corresponding p-values by relying on Sidak's correction \citep[e.g.,  ][Sec 3.4]{kuehl2000}. Specifically, letting $p_r$ be the (smooth test or LRT) p-value obtained for region $r$, the corresponding Sidak's correction is
\begin{equation}
    \label{eqn:sidak}
    p^{Sidak}_r=1-(1-p_r)^R
\end{equation}
with $R$ being the total number of regions being tested.
Notice that the same corrections have also been implemented in Table \ref{table:var-importance2} when assessing the validity of flat backgrounds on the source-free sample. 

\subsubsection{Signal detection via smooth tests}\label{sec:signaldetection}

Similarly to Section \ref{sec:background_only}, we rely on the deviance and the K-statistics (suitably adjusted for post-selection, as described in Section \ref{sec:adjust}) to perform our tests. 

The discovery results on each of the regions of interest are summarised in Table \ref{table:var-importance3}. The second column of the Table \ref{table:var-importance3} corresponds to truncation point, $m$, selected via the $BIC$ criteria in \eqref{eqn:BIC},  when considering a maximum of $M=10$ coefficients. The Sidak adjusted p-values calculated as in \eqref{eqn:sidak} with $R=9$ are reported in parenthesis. Recall that our estimator and test statistics are constructed by sorting the estimated coefficients $\widehat{\theta}_{(j)}$ in \eqref{eqn:coef}. Hence, choosing $m$ to be the point of truncation implies that inference and estimation are performed   considering only the $m$ largest estimated coefficients. 

For regions $\passband{1}$ and  $\passband{2}$, the Bonferroni and K adjusted p-values  are smaller than the global significance level $\alpha=0.05$ even after implementing Sidak's correction. Whereas,   the p-values adjusted for post-selection by means of the naive method, only detect significant deviations over region $\passband{2}$. As discussed in detail in Section \ref{sec:constructUL}, this result is not surprising since the naive approach is the most conservative among the three methods considered. The adjusted p-values for the remaining regions are all equal to one. This implies that our smooth tests analysis allows us to claim that deviations from the background occur only on regions $\passband{1}$ and $\passband{2}$.

The results obtained on the region $\passband{1}$ are consistent with those of \citet{2007ApJ...671..741L}. These are known features, arising in inner accretion disk perhaps, and their detection here is a confirmation that the method is working.

To gain a better understanding of the nature of the deviation from the background model detected on region $\passband{2}$, we rely on the so-called \emph{Comparison Density plot} or \emph{CD-plot} \citep[e.g.,][]{sara2021} shown in Figure \ref{fig:region2}. 
The CD-plot allows us to visualise where the data distribution deviates significantly from the hypothesised distribution (in our case, the re-calibrated background density). It displays the estimated comparison density (dark green solid line) and which, for region $\passband{2}$ specifies as 
\begin{equation}
\begin{aligned}
    \widehat{d}\bigl(u;\widehat{B}_{\passband{2}},F_2\bigl)
    &=   0.9500 + 2.1577 u - 5.2749 u^2 + 2.9179 u^3 
    \label{eqn:d3}
    \end{aligned}
\end{equation}
where $u=\widehat{B}_{\passband{2}}(x)$. Whereas, the green bands are the standard errors of $\widehat{d}\bigl(\widehat{B}_{\passband{2}}(x);\widehat{B}_{\passband{2}},F_2\bigl)$ obtained by simulating from the estimator in (\ref{eqn:d3}) as described in \citet{sara2021}. The grey bands correspond to the $99\%$ confidence bands under the null hypothesis of background only. If the estimated comparison density is within the confidence bands, over the entire range considered, we conclude that there is no significant departure from the background model. Conversely, we expect significant deviations to occur in regions where the estimate lies outside the confidence bands. It is worth emphasising that the CD-plot provides us a representation of the data in the quantile domain; that is, it displays the transformed data $u_i=\widehat{B}_{\passband{2}}(x_i)$, $i=1,\dots,3311$, and their estimated density. Such representation ensures that the most substantial departures of the data distribution from the expected model are magnified and those due to random fluctuations are smoothed out. More details on the construction and discussion of the CD-plot can be found in \citet[][Algorithm 1]{sara2021} and \citet[][Section V A]{algeri2020}.  

For the specific case of Region $\passband{2}$,  the CD-plot in Figure \ref{fig:region2} suggests that significant departures from $\widehat{b}_{\passband{2}}$ occur within the range of $x\in [6.1, 7.9]$\AA. {This detection, however, cannot be attributed to any known spectral features, and it corresponds to the signature of the Chandra optics Iridium absorption edge and is detectable because we are not assuming a particular spectral model here.  
It is worth pointing out that the departure below one at  $x{\gtrsim}$9.1~\AA\ is due to the fact that, since the $ \widehat{d}\bigl(\widehat{B}_{\passband{2}}(x);\widehat{B}_{\passband{2}}, F_2\bigl)$  is the estimate of a density function, its integral is $\approx$1 
(adequate corrections to ensure that the integral of the resulting estimate is exactly $1$ can be implemented as described, for instance, in \cite{sara2021}).
It follows that any peak or departure above $1$ is compensated by a departure below one over the remaining portion of the search region  (in our case, for $x{\gtrsim}$9.1~\AA).

\subsubsection{Signal detection via LRTs} \label{sec:LRTUL}
To further investigate the possibility of undetected spectral lines in regions $W_3$-$W_9$ we proceed by implementing an inferential analysis based on LRTs.

As noted in the introduction, the main advantage of smooth tests is that they allow us to detect deviations from the background occurring at the $m$ ``most significant'' directions identified by the BIC criterion in \eqref{eqn:BIC}.  Therefore, they are completely model-independent, that is, they do not require us to specify the position of the spectral lines or a model for their shape. Nonetheless, when such information is available, one can assess if deviations from the background occur in the direction of the specified signal model by means of the LRT, which is known to be the most powerful test when a model for the signal is provided. 

We begin by assuming that, if a spectral line is present on region $\passband{r}$, its density is\footnote{This is the so-called line response function (LRF) that describes the response of a grating to an infinitesimally narrow line.  It is empirically modeled as a modified Lorentzian, also called a Beta-profile; see Equation~9.1 of the {\sl Chandra} Observatory Proposers' Guide, \url{https://cxc.harvard.edu/proposer/POG/html/chap9.html\#tth_sEc9.3.3}}
\begin{equation}
   \begin{aligned}
s_{r}(x,\widetilde{\mu}_r) &= k(\widetilde{\mu}_r) \Biggl\{1+\biggl[\frac{(x-\widetilde{\mu}_r)}{0.05}\biggl]^2\Biggl\}^{-2.5}
   \end{aligned}    
   \label{eq:LRF}
\end{equation}
with
\begin{equation*}
   \begin{aligned}
\widetilde{\mu}_r &\sim N(\mu_r, 0.005^2)
   \end{aligned}    
\end{equation*}
where $\widetilde{\mu}_{r}$, $r=3,...,9$, is the expected signal location (see column 4 of Table $\ref{table:bands}$) and $k(\widetilde{\mu}_r)$ is the normalising terms depends on the random variable $\widetilde{\mu}_r$. 

We can then specify a suitable model for regions $\passband{3}$-$\passband{9}$ as
\begin{equation}
    \begin{aligned}
q_r(x,\eta_r,\widetilde{\mu}_r) =  (1- \eta_{r}) \widehat{b}_{\passband{r}}(x) + \eta_{r} s_{r}(x,\widetilde{\mu}_r)
    \label{eqn:truef}
    \end{aligned}    
\end{equation}
where $\widehat{b}_{\passband{r}}$ is the background model adequately corrected, as needed, following the method described in Section \ref{sec:4.2}, whereas $\eta_{r}$ is the relative intensity of the expected signal. 
To test if a spectral line with density $s_r$  is present on region $W_r$ we test the hypotheses
\begin{equation}
    \begin{aligned}
    H_0: \eta_r=0 \quad \text{versus} \quad H_1: \eta_r \in (0,1], \quad r=3,...,9
    \label{eqn:hp}
    \end{aligned}
\end{equation}
by means of the test statistic
\begin{equation}
    \begin{aligned}
    \lambda_{r} = -2\sum_{i=1}^n \left\{ \log  \widehat{b}_{W_r}(x_i) -  \mathop{\max}_{\eta_r\in (0,1]}\log q_{r}(x_i,\eta_r,\widetilde{\mu}_r) \right\}.
    \label{eqn:LRT}
    \end{aligned}
\end{equation}
Notice that the parameterisation in \eqref{eqn:truef} implicitly assumes that, under the alternative hypothesis,
the true density, $f_r$, is equal to $q_r$. Therefore, tests based on such model are unable to capture any deviation from $\widehat{b}_{\passband{r}}(x)$ other than the one in the direction of  $s_{r}(x,\widetilde{\mu}_r)$ (hence the analysis in Section \ref{sec:signaldetection}).

Under suitable regularity conditions 
\citep[e.g.,][]{nature}, the null distribution of $\lambda_r$ in \eqref{eqn:LRT} can be approximated by a $\chi^2$ \citep{wilks}. However, since $\eta_r$ in \eqref{eqn:truef} lies on the boundary of its parameter space, under $H_0$, the $\chi^2$ approximation is no longer valid. Nonetheless,  in this setting, the null asymptotic distribution of the LRT is known to be a zero $50\%$ of the times and it is $\chi_1^{2}$ the remaining $50\%$ \citep{chernoff1954}. Therefore, it is sufficient to divide by a factor of two the ``usual'' p-value obtained using the $\chi^2_1$ approximation. 

The results of the LRT for each of the seven regions considered are summarised in Table~$\ref{table:likelihood}$. In the second column, we report the local p-values, that is, the p-values obtained without adjusting for multiple hypothesis testing whereas the global p-values adjusted by means of Sidak's correction in \eqref{eqn:sidak} with $R=7$ are reported in the third column. All the regions apart from  $\passband{6}$ have local p-values larger than $0.05$. 
For region $\passband{6}$, the local p-value $0.0385$. However, once adequately adjusted via Sidak's correction, the p-value for $W_6$ is no longer significant. This result emphasises the importance of implementing adequate corrections when conducting multiple test simultaneously to avoid false discoveries claims.

\subsection{Statistical Properties and Construction of Upper Limits}
\label{sec:constructUL}

In order to confirm the validity of the analyses performed in Sections \ref{sec:signaldetection}-\ref{sec:LRTUL}, it is important to investigate the statistical properties (power and probability of type I error) for each of the inferential procedure considered. We will then exploit the power curves obtained to construct suitable upper limits as prescribed in \citet{2010ApJ...719..900K}.

We investigate the power of the proposed statistical tests by simulating  Monte Carlo samples from $q_r$ in \eqref{eqn:truef} for different intensity levels $\eta_r$. The size of each simulated dataset is the same as that of the original sample observed on the seven regions considered (i.e., $n_{r}$ with $r=3,\dots,9$ in Table \ref{table:bands}). Using the simulated data, we implement the methods described  in Section $\ref{s:methods}$ as well as the LRT procedure discussed  in Section \ref{sec:LRTUL}. When the true value of $\eta_r$ is non-zero,  the null hypothesis in \eqref{eqn:hp} is correctly rejected if the p-value is smaller than the pre-specified significance level. We repeat this process $5000$ times, and at each replicate, we record the number of rejections of the null. The proportion of rejections corresponds to the simulated power for the (non-zero) values of $\eta_r$ fixed when conducting the simulation. We repeat the simulation for different values of $\eta_r$ in order to generate a series of power curves, one for each region considered. The latter is displayed in Figure \ref{fig:compare_fig}. The same analysis is also repeated by adjusting the respective p-values for multiple hypothesis testing via Sidak's correction in \eqref{eqn:sidak} at each replicated of the simulation. As expected, adjusting for multiple hypothesis testing leads to a reduction of the power. Not surprisingly, the power of the LRT is higher than any other procedure considered. That is because, smooth tests do not require the specification of the signal model and simply assess for the presence of deviations from the background.   

The probabilities of type I error for each of the procedures considered (obtained by setting $\eta_r=0$)   are reported in Tables \ref{table:var-importance}-\ref{table:var-importance_sidak}. The simulated type I error using Bonferroni adjustment for the deviance statistic often exceeds the  significance level of $0.05$ in the context of local analysis. It also exceeds the significance level $\alpha_r=0.0073$ in the global analysis; that is when setting the probability of false discovery across all the seven regions to be $\alpha=0.05$. Conversely, the K-statistic and the naive post-selection adjustment for the deviance perform well in controlling the specified significance level even though the naive adjustment appears to be excessively conservative (the respective probability of type I error is always zero in the global analysis). 
This is reflected also in the power curves reported in Figure~\ref{fig:compare_fig}. The naive approach is the most conservative among the three, whereas Bonferroni exhibits the highest power.

Finally,  upper limits are constructed by inverting the power curves of the LRT at $50\%$ and $90\%$ and multiplying the resulting $\eta_r$ value for the sample size, $n_r$ (see Table \ref{table:bands}) of the discovery region $\passband{r}$, for $r=3,\dots,9$. 
The results are summarised in Table~\ref{table:UL_LRT}. We can interpret the Sidak adjusted upper limits as the number of samples from the expected signal needed to achieve the specified power when testing simultaneously regions $\passband{3}$-$\passband{9}$. For example, for region $\passband{3}$, our $50\%$ upper limit computed using the LRT after Sidak correction is 40 (39.42). This tells us that if a spectral line at position 12.131  \AA \ was present, we would need 40 events  in this location (out of the 730 observed in the entire $\passband{3}$ region) to be able to detect such spectral line with power 50\%, while simultaneously looking for spectral lines in the regions $\passband{4},\dots,\passband{9}$.
Whereas, if we were interested in designing a future observation targeting solely region $\passband{3}$, our $50\%$ upper limit computed using the (local) LRT is 30 (29.93). This tells us that if a spectral line at position 12.131  \AA \ was present, we would need 30 events at such location to detect it with power 50\%, and assuming that no other test on other regions is conducted at the same time.
Similar interpretations can be given to the $90\%$ upper limits and for other regions. 

For the sake of comparison, the upper limits obtained by means of smooth tests are reported in Tables \ref{table:vsd}-\ref{table:vsd2}. Not surprisingly, since smooth tests do not rely on the specification of a model for the signal, they are more conservative than the LRT. For example, for  region $\passband{3}$, the $50\%$ upper limits 
computed using the Bonferroni, K-statistic and the naive methods, and adjusted via Sidak for multiple hypothesis testing lead to 53, 64, and 68 events, respectively.

\begin{figure*}
\centering

\subfigure{
    \begin{minipage}[t]{0.5\linewidth}
        \centering
       \includegraphics[width=3.525in]{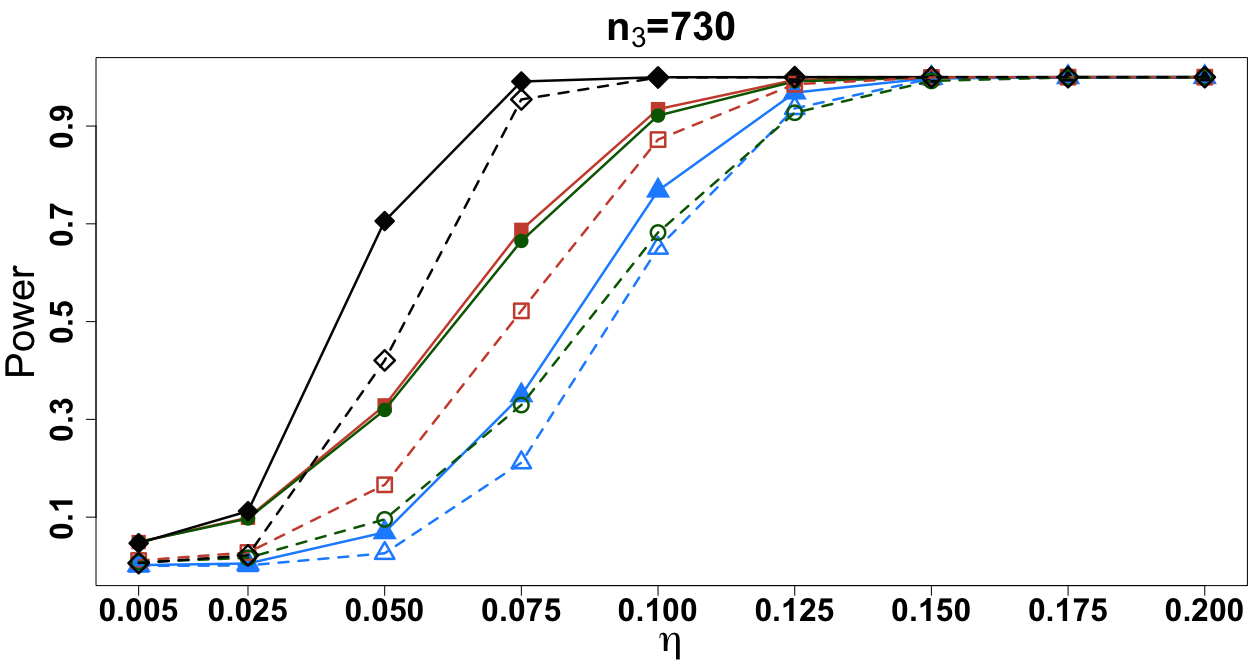}\\
       \vspace{0.15cm}
        \includegraphics[width=3.525in]{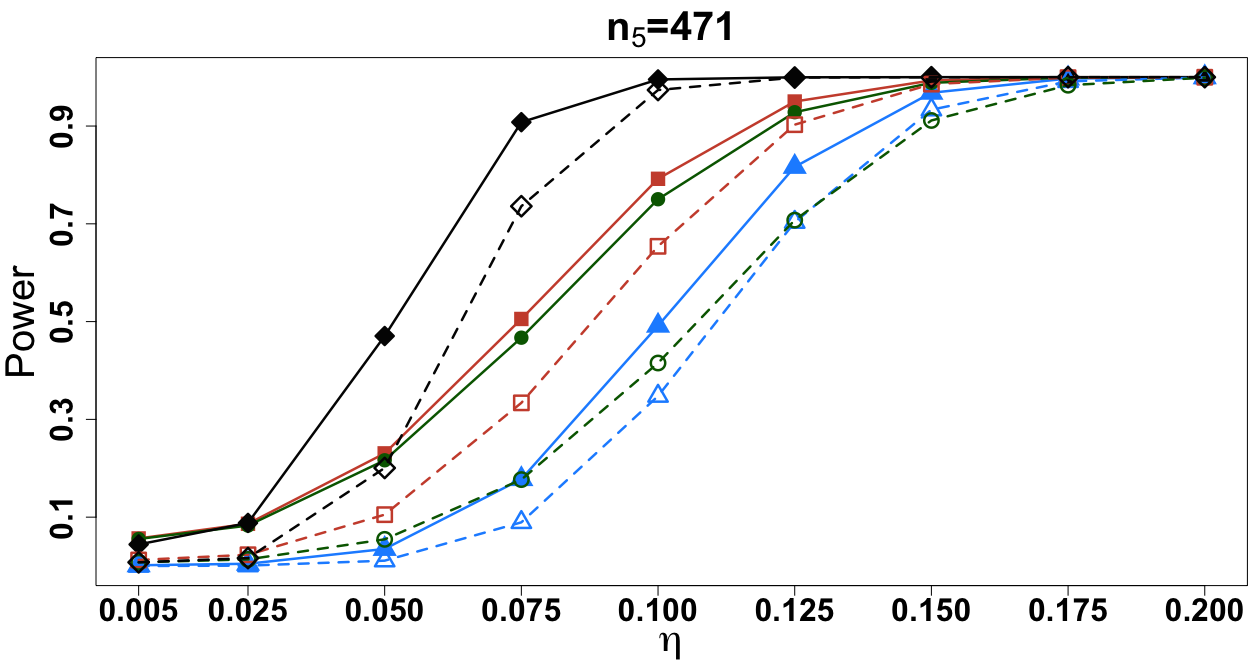}\\
        \vspace{0.15cm}
        \includegraphics[width=3.525in]{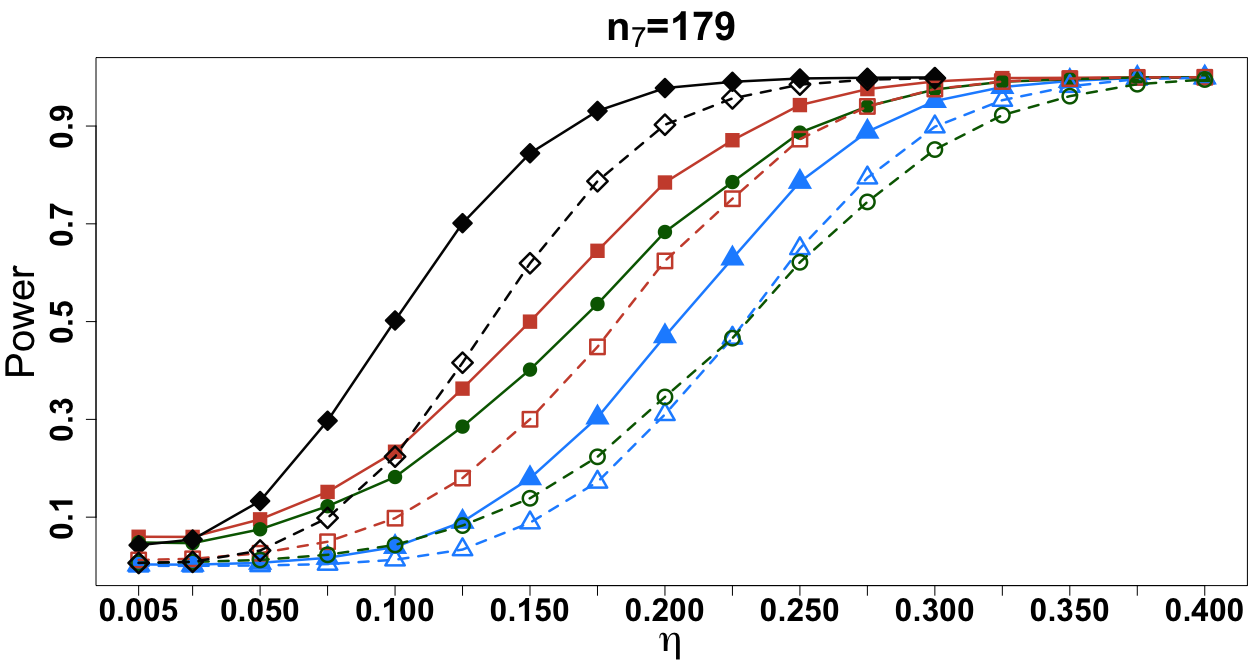}\\
        \vspace{0.15cm}
        \includegraphics[width=3.525in]{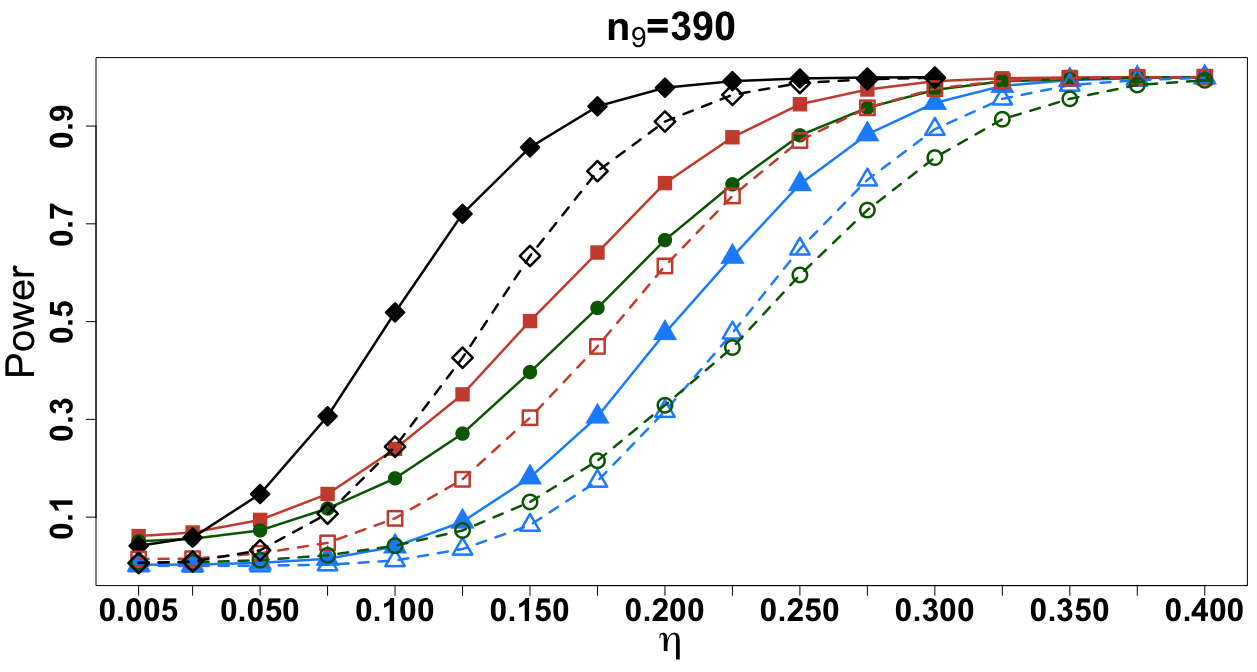}\\
        \vspace{0.15cm}
    \end{minipage}%
}%
\subfigure{
    \begin{minipage}[t]{0.5\linewidth}
        \centering
        \includegraphics[width=3.525in]{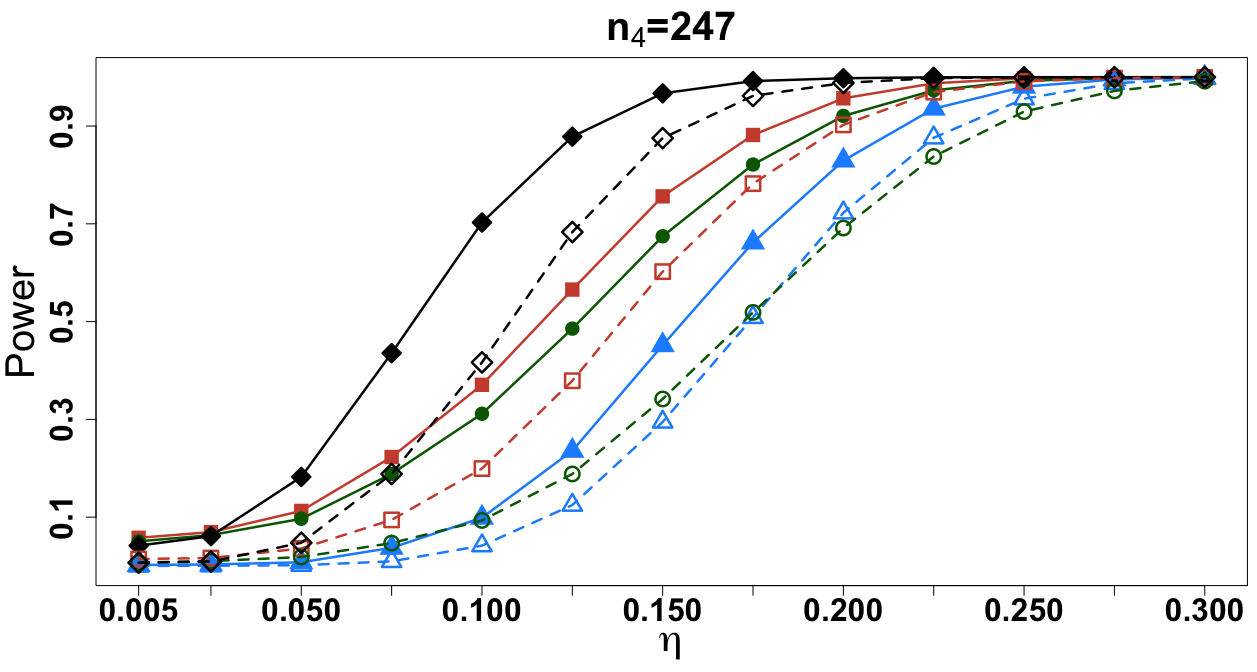}\\
        \vspace{0.15cm}
        \includegraphics[width=3.525in]{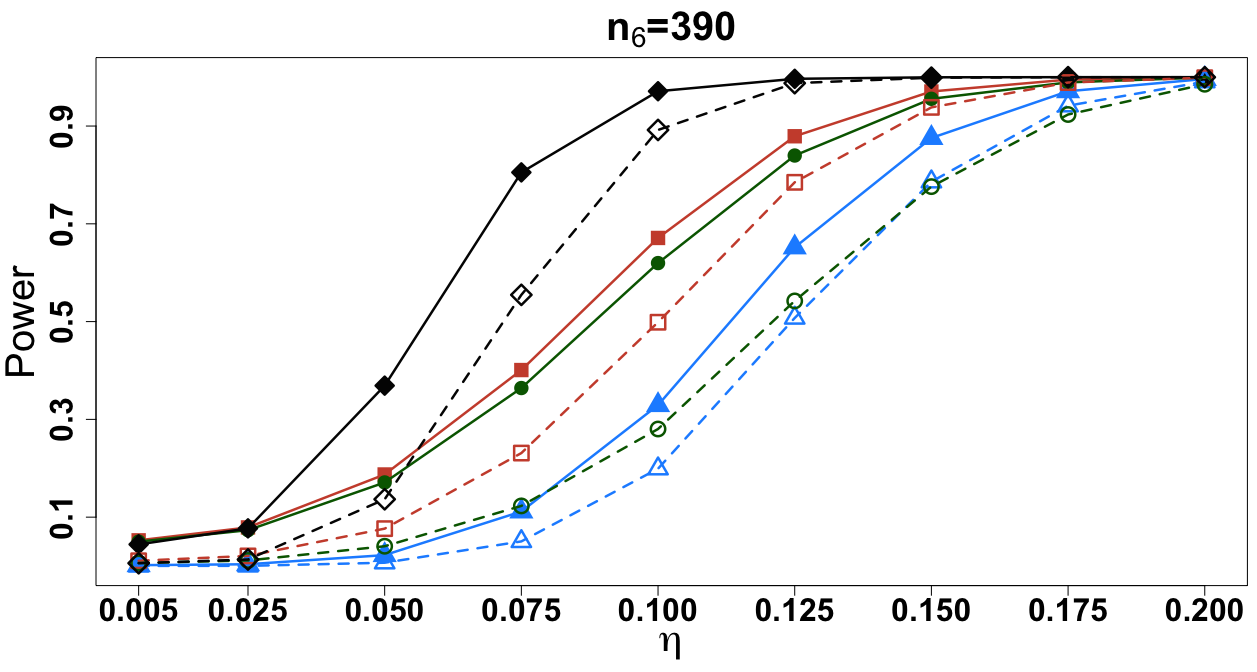}\\
        \vspace{0.15cm}
        \includegraphics[width=3.525in]{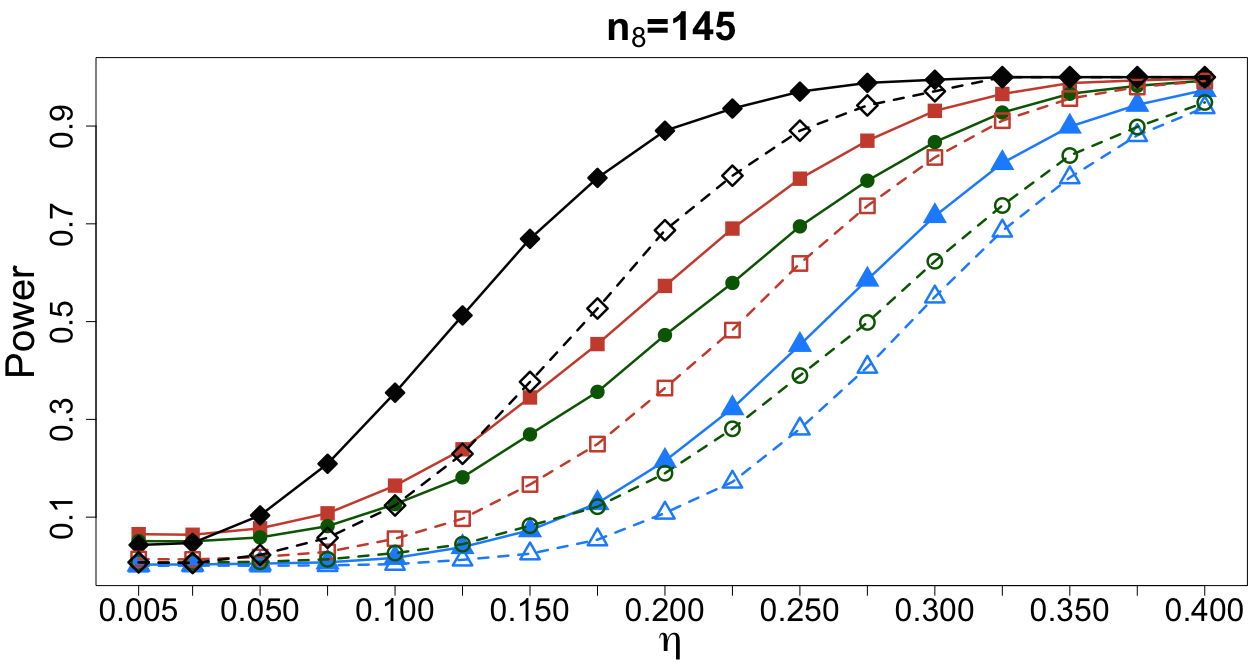}\\
        \vspace{0.15cm}
        \includegraphics[width=3.525in]{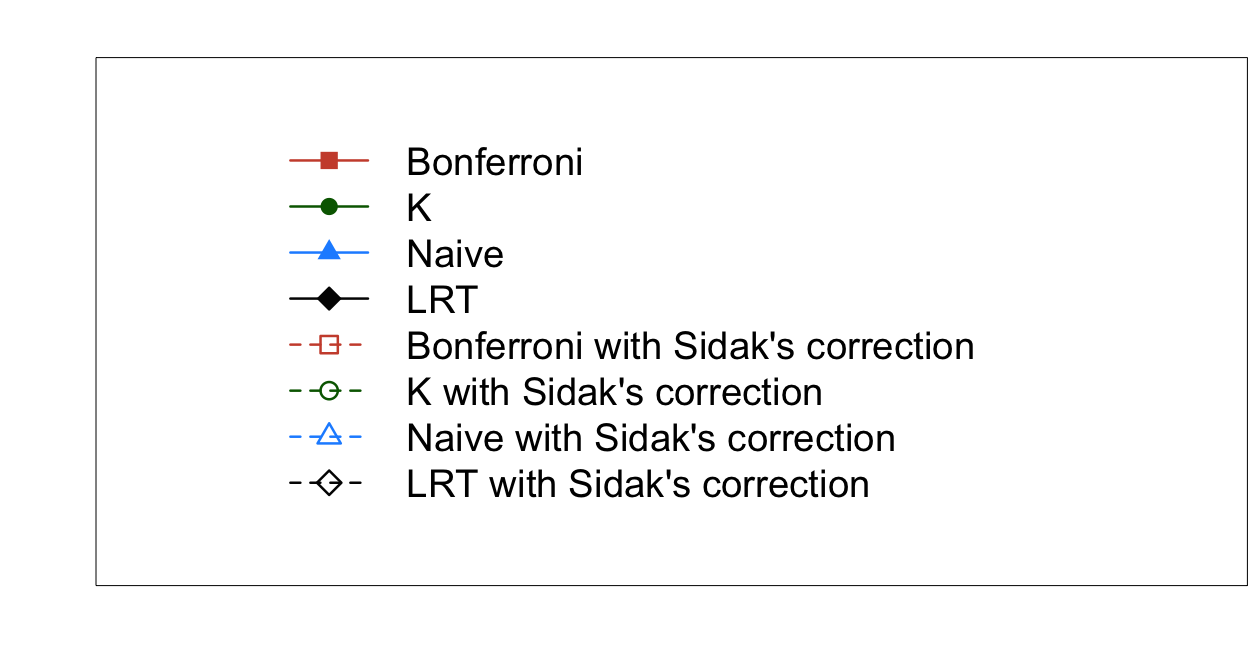}\\
        \vspace{0.15cm}
    \end{minipage}%
}%
\caption{Simulated power curves for regions $\passband{3}$-$\passband{9}$. Red-square  and blue-triangle lines correspond to the power curves for the deviance adjusted for post-selection via Bonferroni and the naive approach, respectively. The green-circle lines are the power curves for the K-statistic suitably adjusted for post-selection. The black-diamond lines are the power curves for the LRT. 
The solid lines correspond to the local power curves, with $\alpha_r=0.05$, for each $r=3,\dots,9$. The dashed lines are the power curves adjusted for multiple testings using Sidak's correction to ensure a global significance of $\alpha=0.05$ (local significance  $\alpha_r=0.0073$, for each $r=3,\dots,9$). }
\centering
\vspace{0.2cm}
\label{fig:compare_fig}
\end{figure*}

\begin{table*}
	\centering
	\begin{tabular}{
	|c!{\vrule width 1.4pt}
	cc!{\vrule width 1.4pt}
	cc!{\vrule width 1.4pt}}
		\hline
		\multirow{3}{*}{Regions ($\passband{r}$)} & 
		\multicolumn{2}{c!{\vrule width 1.4pt}}{50$\%$ upper limits via LRT} & \multicolumn{2}{c!{\vrule width 1.4pt}}{90$\%$ upper limits via LRT} \\
		\cline{2-5}
	&Local & Sidak adjusted&Local & Sidak adjusted\\
		\hline        
	    $\passband{3}$&29.93& 39.42   & 48.91 &53.29\\
		$\passband{4}$ &20.00 & 26.43 &32.36 &39.52 \\
		$\passband{5}$&24.02 & 30.14&35.32& 43.80\\
		$\passband{6}$&22.62 &28.08 & 34.71 & 39.39\\
		$\passband{7}$ & 17.90  & 24.17  &29.71 & 35.98\\
		$\passband{8}$ & 17.84&24.80 & 30.30 &36.25 \\
		$\passband{9}$ &  37.83& 21.87  &63.57& 76.83\\
		\hline
	\end{tabular}
      \caption{$50\%$ and $90\%$ upper limits on different regions using the LRT, with and without Sidak's correction. The $50\%$ upper limits are calculated by the proportion of the expected lines $\eta$ to achieve the power 0.5 (solid line in Figure \ref{fig:compare_fig}) times the sample sizes $n_r$. Similarly, the $90\%$ upper limits are calculated by the proportion of the expected lines $\eta$ to achieve the power 0.9 times the sample sizes $n_r$.}
	\label{table:UL_LRT}
\end{table*}

\begin{table*}
    \centering
	\begin{tabular}{
	|c!{\vrule width 1.4pt}
	c|c|c|c|c|c|c!{\vrule width 1.4pt}}
		\hline
	 & Shifted & Local &Sidak's  & LRT 50\% & LRT 90\% & Sidak's LRT 50\%& Sidak's LRT 90\% \\
		& lines [\AA] & p-values&correction & Upper limits & Upper limits & Upper limits &Upper limits \\
		\hline 
	    $\passband{4}$ & 14.908 &0.5000 &0.9921& 20.00& 32.60&  26.92& 39.77\\
		$\passband{5}$ & 16.930 &0.0017 &0.0118& 24.54& 34.85&  30.14& 43.80\\
		\hline
	\end{tabular}
      \caption{Local $p$-values for the LRT and adequate multiple hypothesis testing adjustments when investigating the presence of the  16.93~\AA\  emission line in $\passband{5}$.}
	\label{table:likelihood5}
\end{table*}

\section{Discussion}\label{sec:discuss}

\subsection{Advantages and Limitations}

We have developed a novel method to detect weak signals distinct from a smooth background in high-resolution photon counting spectra.  This approach anticipates difficulties likely to be encountered in the coming era of calorimeter spectra. The method is implemented to work with unbinned photon lists that allows the full available spectral resolution to be used, though a modification to use binned spectra is viable from an algorithmic perspective and it is the subject of future work.  

The statistical methodology presented here is particularly advantageous at high resolution because a precise specification of the source model spectrum is often not possible as the information available in the data usually exceeds that in the models proposed. Here we show that one can indeed exploit this phenomenon by modeling and estimating the ``gap'' between the  (potentially misspecified) model available and the true spectrum using smooth functions like shifted Legendre polynomials. On this note, it is worth emphasizing that, as proven in \citet{algeri2020}, the closer the postulated model is to the truth, the more accurate (less biased) is the estimate of the latter. It follows that, in principle, one could avoid specifying a model for the spectrum and estimate it by means of  smooth functions. Nonetheless, if a model is available (even if misspecified), it should be used in order  to reduce the gap between the proposed model and the truth. 

The implementation currently ignores spectral calibration products like the effective area and the redistribution matrices, and therefore cannot be applied to CCD resolution spectra.  Furthermore, the method relies on a comparison between the smooth model description of the source-free background and the source+background datasets, so it cannot be applied to cases where the background is contaminated by the source or where the background is not smoothly varying.

\subsection{Inferences based on \rtcru\ analysis}

\subsubsection{Domain of applicability}

We first note that our method easily detects the presence of significant source emission in passband $\passband{1}$.  This is not surprising, as these lines have been identified and analyzed by several studies (e.g., \citet{2007ApJ...671..741L} resolved it clearly in HETGS+ACIS-S spectra; and \citet{2021MNRAS.500.4801D} successfully modeled the triplet in the same dataset that we use).  The chance that a random fluctuation can produce a detectable departure from the background is assessed as $p{\lesssim}10^{-2}$ after accounting for multiple hypothesis tests (see Table~\ref{table:var-importance3}).  This serves as a validation of the method, in that a line complex known to exist is correctly found.

An important characteristic of our method is that it is not limited to narrow lines.  If the source spectrum has a different shape compared to the background spectrum, we expect that to be detected, i.e., the method allows for a differentiation between the continuum and the background.  This was achieved serendipitously in passband $\passband{2}$, where the existence of a feature is flagged with high significance ($p{\ll}10^{-10}$; see Table~\ref{table:var-importance3}).  The detected feature (see Figure~\ref{fig:fullbands}) is a characteristic of the response of the \chandra\ mirror coating, exhibiting an edge due to the absorption of incident photons by Iridium.  Such features are usually not visible in residuals in standard spectral analyses because the modeling directly incorporates the sensitivity of the telescope system and suitably weights the incident photon spectrum.  Because our method does not include such calibration products, it can be used to locate and study such features, incorporating known calibration (as is done for line emission in Equations~\ref{eq:LRF}, \ref{eqn:truef}), thus potentially providing independent measures of uncertainty on the calibration.

\subsubsection{Thermal line emission}

We have explored the possibility of detecting emission from some prominent soft X-ray emission lines (Table~\ref{table:bands}; see also Section~\ref{s:data}) and find that in no case are any of the lines we searched for detectable.  Here we compare the model predicted counts obtained using the spectral model of \citet{2021MNRAS.500.4801D} with a nominal estimate of background corrected counts estimate obtained via a Bayesian estimate (as in \citealt{2002ApJS..138..185F}), with the 50\% upper limits obtained using smooth tests and post-selection adjustments (Section~\ref{s:analysis}; Table~\ref{table:UL_LRT}):
\begin{itemize}
    \item[] $\passband{3}$ : This passband is centered on the Ne\,X resonance line at 12.14~\AA.  The model predicts $\approx$16~counts in this region, and a measurement of the counts over the width of the LRF yields an estimate of $\approx$20~counts, with a 68\% uncertainty interval of $\approx\pm$10.  This would normally be considered a good match between model and data, but the upper limit at this location is $\sim$30~counts, increasing to $\sim$40~counts when corrections due to multiple hypotheses testing are included, and we conclude that the line is not detectable.  This assessment is supported by a visual inspection of the observed counts spectrum, where an  enhancement in counts at the location of the line is not apparent.
    \item[] $\passband{4}$ and $\passband{5}$ : The dominant features in these bands are expected to be from the Fe\,XVII lines at 15.014~\AA, 17.051~\AA\ and 17.096~\AA.  Note that these lines have peak emissivity at $\approx$5~MK, whereas the best-fit spectral model suggests a plasma temperature of $\approx$14~MK.  The emissivities at such high temperatures are reduced by a factor of 30, but they remain the strongest lines from a thermal spectrum and have approximately similar intensities.  The predicted counts from the best-fit model is $\approx$10~counts in each, and the measured values are $\approx$20 and $\approx$6 counts in $\passband{4}$ and $\passband{5}$ respectively.  These estimates are prima facie inconsistent with theoretical expectations, though the posterior distributions of the intensities do overlap.  The local upper limits are 20 and 24 counts, increasing to 26 and 30 counts at the 50\% level (see Table~\ref{table:UL_LRT}) after accounting for multiple hypotheses, for $\passband{4}$ and $\passband{5}$ respectively.  This suggests that the 15~\AA\ line is a borderline detectable feature, but the 17~\AA\ lines are not detected.

    However, notice that in the counts spectrum in Figure~\ref{fig:fullbands}, there is a relatively strong feature at $\approx$16.93~\AA, about 2100~km~s$^{-1}$ blueward of 17.051~\AA\ (marked with a brown cross-hatch).  
    We find that the feature is detected with a significance of $p{\lesssim}0.002$ locally and at $p{\lesssim}0.012$ after multiple hypothesis correction (see Table~\ref{table:likelihood5}), with a 50\% upper limit $<25$~counts.  The estimated signal strength for this feature is 16.9~$^{<26.9}_{>8.5}$~counts (the bounds represent 68\% HPD uncertainty interval as in Table~\ref{tab:modelct}), consistent with the putative upper limit.  Inner accretion disk velocities of $\sim$10$^3$~km~s$^{-1}$ are plausible for symbiotic variables \citep[see, e.g.,][]{1985ApJ...297..275R}, though the required shift here is twice as much.  A hot spot that can achieve such a large blue shift would be located at a height of $\approx$0.06~R$_{\odot}$ above a 1.4~M$_{\odot}$ white dwarf, which is not infeasible.
    But if this were a blue-shifted Fe\,XVII~$\lambda$17.051 line, we should expect its 15~\AA\ counterpart to also be similarly blue shifted.  The location of this putative blue-shifted 15~\AA\ line is also marked with a brown cross-hatch in the $\passband{4}$ panel in Figure~\ref{fig:fullbands}.  We find no detectable feature at this location, with a multiple-hypothesis corrected 50\% upper limit at $<27$~counts (Table~\ref{table:likelihood5}; the estimated counts are 0~$^{<8.2}_{\ge0}$).  We thus conclude that the feature at 16.93~\AA\ is {\sl not} a blue-shifted thermal line.  We discuss its possible origin further in Section~\ref{sec:photorefl} below.
    \item[] $\passband{6}$ : The strongest feature in this passband is the O\,VIII resonance line at 18.96~\AA.  As can be seen in Figure~\ref{fig:fullbands}, there is an apparent enhancement at where the line is expected to be, though the predicted model counts are more than $2\times$ smaller compared to the estimated counts, and indeed the predicted counts value falls outside the 68\% uncertainty bounds (Table~\ref{tab:modelct}).  Interestingly, the local upper limit shows that the deviation is significant with $p<0.05$, but when corrections due to multiple hypothesis tests are included, there is no significant evidence for a detection.  The 50\% upper limit is $<28$, and we conclude that the data are not of suitable quality to constrain the Oxygen emission.
    \item[] $\passband{7}$-$\passband{9}$ : These three bands include the He-like density sensitive triplet of O\,VII at 21.6, 21.8, and 22.1~\AA.  We find no evidence for a detection of any of the lines, with the corrected $p{\approx}1$ in all three cases, and with 50\% upper limits in the $21-24$~counts range.  However, this case provides a cautionary illustration of the pitfalls of ignoring detectability.  The nominal estimates of the counts in each of the lines shows that the intercombination line at 21.8~\AA\ has a brightness of $\approx$10 counts, higher than both the resonance and forbidden line brightnesses.  Without the upper limits analysis demonstrating that none of the lines are detectable, it is easy to over-interpret the high brightness estimate of the intercombination line as indicative of a very high density ($\gg$10$^{13}$~cm$^{-3}$; see, e.g., \citealt{2001ApJ...556L..91S}) in the emitting plasma.  While the presence of such high densities cannot be formally excluded, the current dataset is of insufficient quality to place a definitive constraint.  Observations with new observatories like XRISM, Athena, or LEM are necessary to resolve this question.
\end{itemize}

\subsubsection{The Origin of the 16.93~\AA\ Feature}\label{sec:photorefl}

The hard emission in the  6-7 keV range likely originates close to the accretor. The 6.4 keV Fe Kalpha line could originate from a region of the boundary layer of the accretion disk or in a bright spot on the disk; The 6.7 keV and the 7 keV, can be attributed to He-like and H-like iron emission lines, and could be produced via photoionization and collisional ionization/excitation mechanisms in the hot plasma, either in the vicinity of the WD or the inner-jet region, or both  (see, e.g., \citealt{2009ApJ...701.1992K}, \citealt{2010ApJ...710L.132K}, \citealt{2014MNRAS.437..857E}).  We speculate that the same photoionization or reflection spectrum photon reprocessing could be the source of the line feature detected at 16.93~\AA.  Several plausible candidate ionic species ranging from Fe\,XIII to Fe\,XIX exist in the XSTAR \citep{1999ascl.soft10008K} line list within $\approx$500~km~s$^{-1}$ of the detected line.  The reflection spectrum model {\tt xillver} \citep{2013ApJ...768..146G} shows the existence of emission localized at 16.93~\AA\ for flat incident spectra ($\Gamma\approx$1-1.2), moderate Fe abundances (A$_{\rm Fe}\approx$0.5-1), ionization (log$\xi\approx$2-3). Intriguingly, a feature at the same location was found by \citet{2021ApJ...920..142H} in the \chandra\ LETG spectrum of the ultracompact X-ray binary 4U\,1626–67; they are also unable to identify the line, and suggest a photoionized emission as the origin.  Unlike the double-peaked emission that \citeauthor{2021ApJ...920..142H} find that is reminiscent of an accretion-disk origin, the profile of the line in \rtcru\ is sharply single-peaked, suggesting an origin in a hot spot or an inner jet.
A detailed study of the origin of this line, and a search for other such lines is beyond the scope of both this article and the \chandra\ dataset; we expect that observations obtained with calorimeter resolutions will confirm the presence of such lines and lead to a better understanding of the emission from \rtcru\ and other symbiotics.

\section{Summary}\label{sec:summary}

We present a method designed to correctly analyze high-resolution spectra affected by high background. The method is currently implemented for photon lists, but extensions to binned spectra such as those obtained by XMM/RGS are possible in principle and are the subject of future work.  We first characterize the background and determine whether the background-contaminated source spectrum differs significantly in its shape over any passband.  This allows us to detect both line emission as well as continuum variations.  We properly correct for multiple hypothesis tests, and where a spectral line is not detected, we place upper limits on the source intensity.  From a statistical perspective, in addition to the novel analytical framework presented here, we introduce a new test statistic to perform smooth tests and adequate post-selection inferential adjustments. The latter are obtained by simply inverting the power function of our post-selection adjusted smooth tests and obtained via Monte Carlo simulations.

We apply this method to the \chandra\ LETGS+HRC-S spectrum of the symbiotic star \rtcru.  Because of the high background, traditional techniques designed to detect and identify lines and compute fluxes are unfeasible for these data.  Consequently, a global fit was carried out by \citet{2021MNRAS.500.4801D} using a power-law and a thermal component, supplemented by Gaussian lines to model the Fe~K$\alpha$ region.  We first validate our method by confirming the detection of the Fe-line triplet in the 6-7~keV region, and further demonstrate the ability of the method to find variations in the continuum spectrum by detecting the Iridium M edge present due to the mirror coating in all \chandra\ observations.

Since the clear detection of low energy spectral lines would be an important diagnostic for the emission mechanisms that operate in symbiotic systems, we have employed smooth models and smooth tests to correct, when needed, the uncertain background model and to determine whether source spectra show distinct differences in shape compared to the background in several wavelength bands where spectral lines are expected to be present.  None of the expected thermal lines are detected, and the upper limits are all larger than the predicted model counts.  With the analysis of the O\,VII He-like triplet, we illustrate how checking for detectability allows us to avoid mistakenly claiming that the emitting plasma is at high density.  We thus conclude that the fitted model considered by \citet{2021MNRAS.500.4801D} is adequate in its overall characteristics to explain the emission from \rtcru.

We serendipitously and unambiguously detect emission in a line-like feature located at 16.93~\AA \ (0.732 keV), and conclude that it cannot be attributed to thermal plasma emission.  We speculate that it is derived from a hitherto unmodeled photoionization of reflection spectrum component.  Future observations with missions like XRISM \citep{2022SPIE12181E..1SI}, Athena \citep{2022arXiv220814562B}, etc., using calorimeter-level spectral resolutions and high effective areas are necessary to detect and model such processes.

\section*{Acknowledgements}
Sara Algeri and Xiangyu Zhang are grateful for the financial support provided by the Office of the Vice President for Research at the University of Minnesota.
Vinay Kashyap was supported by the NASA Contract NAS8-03060 to the \chandra\ X-ray Center.  Margarita Karovska and Vinay Kashyap acknowledge support for this work provided via the \chandra\ grant GO5-16023X.

\section*{Data and code Availability}
The data and codes used for the analyses in Section \ref{s:analysis} are available at 
the github site at \url{http://github.com/xiangyu2022/Symbiotic-Star-RT-Cru-Analysis}.

\nocite*{}  
\bibliographystyle{mnras}
\bibliography{bibfile.bib} 

\begin{thebibliography}{}
\makeatletter
\relax
\def\mn@urlcharsother{\let\do\@makeother \do\$\do\&\do\#\do\^\do\_\do\%\do\~}
\def\mn@doi{\begingroup\mn@urlcharsother \@ifnextchar [ {\mn@doi@}
  {\mn@doi@[]}}
\def\mn@doi@[#1]#2{\def\@tempa{#1}\ifx\@tempa\@empty \href
  {http://dx.doi.org/#2} {doi:#2}\else \href {http://dx.doi.org/#2} {#1}\fi
  \endgroup}
\def\mn@eprint#1#2{\mn@eprint@#1:#2::\@nil}
\def\mn@eprint@arXiv#1{\href {http://arxiv.org/abs/#1} {{\tt arXiv:#1}}}
\def\mn@eprint@dblp#1{\href {http://dblp.uni-trier.de/rec/bibtex/#1.xml}
  {dblp:#1}}
\def\mn@eprint@#1:#2:#3:#4\@nil{\def\@tempa {#1}\def\@tempb {#2}\def\@tempc
  {#3}\ifx \@tempc \@empty \let \@tempc \@tempb \let \@tempb \@tempa \fi \ifx
  \@tempb \@empty \def\@tempb {arXiv}\fi \@ifundefined
  {mn@eprint@\@tempb}{\@tempb:\@tempc}{\expandafter \expandafter \csname
  mn@eprint@\@tempb\endcsname \expandafter{\@tempc}}}

\bibitem[\protect\citeauthoryear{Algeri}{Algeri}{2020}]{algeri2020}
Algeri S.,  2020, Phys. Rev. D, 101, 015003

\bibitem[\protect\citeauthoryear{Algeri}{Algeri}{2021}]{algeri2021}
Algeri S.,  2021, Electronic Journal of Statistics, pp 5570--5597

\bibitem[\protect\citeauthoryear{Algeri \& Zhang}{Algeri \&
  Zhang}{2021}]{sara2021}
Algeri S.,  Zhang X.,  2021, Journal of Computational and Graphical Statistics,
  pp 1--12

\bibitem[\protect\citeauthoryear{Algeri, van Dyk, Conrad  \& Anderson}{Algeri
  et~al.}{2016}]{algeri2016}
Algeri S.,  van Dyk D.~A.,  Conrad J.,   Anderson B.,  2016, Journal of
  Instrumentation, 11, P12010

\bibitem[\protect\citeauthoryear{Algeri, Aalbers, Mor{\aa}  \& Conrad}{Algeri
  et~al.}{2020}]{nature}
Algeri S.,  Aalbers J.,  Mor{\aa} K.~D.,   Conrad J.,  2020, Nature Reviews
  Physics, 2, 245

\bibitem[\protect\citeauthoryear{{Barret} et~al.,}{{Barret}
  et~al.}{2022}]{2022arXiv220814562B}
{Barret} D.,  et~al., 2022, arXiv e-prints, \href
  {https://ui.adsabs.harvard.edu/abs/2022arXiv220814562B} {p. arXiv:2208.14562}

\bibitem[\protect\citeauthoryear{{Bird} et~al.,}{{Bird}
  et~al.}{2007}]{2007ApJS..170..175B}
{Bird} A.~J.,  et~al., 2007, \mn@doi [\apjs] {10.1086/513148}, \href
  {https://ui.adsabs.harvard.edu/abs/2007ApJS..170..175B} {170, 175}

\bibitem[\protect\citeauthoryear{Chernoff}{Chernoff}{1954}]{chernoff1954}
Chernoff H.,  1954, \mn@doi [The Annals of Mathematical Statistics]
  {10.1214/aoms/1177728725}, 25, 573

\bibitem[\protect\citeauthoryear{{Cieslinski}, {Elizalde}  \&
  {Steiner}}{{Cieslinski} et~al.}{1994}]{1994A&AS..106..243C}
{Cieslinski} D.,  {Elizalde} F.,   {Steiner} J.~E.,  1994, \aaps, \href
  {https://ui.adsabs.harvard.edu/abs/1994A&AS..106..243C} {106, 243}

\bibitem[\protect\citeauthoryear{Cox}{Cox}{1975}]{cox1975}
Cox D.,  1975, Biometrika, 62, 441–444

\bibitem[\protect\citeauthoryear{{Danehkar}, {Karovska}, {Drake}  \&
  {Kashyap}}{{Danehkar} et~al.}{2021}]{2021MNRAS.500.4801D}
{Danehkar} A.,  {Karovska} M.,  {Drake} J.~J.,   {Kashyap} V.~L.,  2021,
  \mn@doi [\mnras] {10.1093/mnras/staa3554}, \href
  {https://ui.adsabs.harvard.edu/abs/2021MNRAS.500.4801D} {500, 4801}

\bibitem[\protect\citeauthoryear{{Ducci}, {Doroshenko}, {Suleimanov},
  {Niko{\l}ajuk}, {Santangelo}  \& {Ferrigno}}{{Ducci}
  et~al.}{2016}]{2016A&A...592A..58D}
{Ducci} L.,  {Doroshenko} V.,  {Suleimanov} V.,  {Niko{\l}ajuk} M.,
  {Santangelo} A.,   {Ferrigno} C.,  2016, \mn@doi [\aap]
  {10.1051/0004-6361/201628242}, \href
  {https://ui.adsabs.harvard.edu/abs/2016A&A...592A..58D} {592, A58}

\bibitem[\protect\citeauthoryear{{Eze}}{{Eze}}{2014}]{2014MNRAS.437..857E}
{Eze} R.~N.~C.,  2014, \mn@doi [\mnras] {10.1093/mnras/stt1947}, \href
  {https://ui.adsabs.harvard.edu/abs/2014MNRAS.437..857E} {437, 857}

\bibitem[\protect\citeauthoryear{{Freeman}, {Kashyap}, {Rosner}  \&
  {Lamb}}{{Freeman} et~al.}{2002}]{2002ApJS..138..185F}
{Freeman} P.~E.,  {Kashyap} V.,  {Rosner} R.,   {Lamb} D.~Q.,  2002, \mn@doi
  [\apjs] {10.1086/324017}, \href
  {https://ui.adsabs.harvard.edu/abs/2002ApJS..138..185F} {138, 185}

\bibitem[\protect\citeauthoryear{{Garc{\'\i}a}, {Dauser}, {Reynolds},
  {Kallman}, {McClintock}, {Wilms}  \& {Eikmann}}{{Garc{\'\i}a}
  et~al.}{2013}]{2013ApJ...768..146G}
{Garc{\'\i}a} J.,  {Dauser} T.,  {Reynolds} C.~S.,  {Kallman} T.~R.,
  {McClintock} J.~E.,  {Wilms} J.,   {Eikmann} W.,  2013, \mn@doi [\apj]
  {10.1088/0004-637X/768/2/146}, \href
  {https://ui.adsabs.harvard.edu/abs/2013ApJ...768..146G} {768, 146}

\bibitem[\protect\citeauthoryear{{Hemphill}, {Schulz}, {Marshall}  \&
  {Chakrabarty}}{{Hemphill} et~al.}{2021}]{2021ApJ...920..142H}
{Hemphill} P.~B.,  {Schulz} N.~S.,  {Marshall} H.~L.,   {Chakrabarty} D.,
  2021, \mn@doi [\apj] {10.3847/1538-4357/ac0ade}, \href
  {https://ui.adsabs.harvard.edu/abs/2021ApJ...920..142H} {920, 142}

\bibitem[\protect\citeauthoryear{{Ishisaki} et~al.,}{{Ishisaki}
  et~al.}{2022}]{2022SPIE12181E..1SI}
{Ishisaki} Y.,  et~al., 2022, in {den Herder} J.-W.~A.,  {Nikzad} S.,
  {Nakazawa} K.,  eds,  Society of Photo-Optical Instrumentation Engineers
  (SPIE) Conference Series Vol. 12181, Society of Photo-Optical Instrumentation
  Engineers (SPIE) Conference Series. p. 121811S, \mn@doi{10.1117/12.2630654}

\bibitem[\protect\citeauthoryear{Kallenberg \& Ledwina}{Kallenberg \&
  Ledwina}{1997}]{Led1997}
Kallenberg W. C.~M.,  Ledwina T.,  1997, Journal of the American Statistical
  Association, 92, 1094

\bibitem[\protect\citeauthoryear{{Kallman}}{{Kallman}}{1999}]{1999ascl.soft10008K}
{Kallman} T.,  1999, {XSTAR: A program for calculating conditions and spectra
  of photoionized gases}, Astrophysics Source Code Library, record
  ascl:9910.008 (\mn@eprint {ascl} {9910.008})

\bibitem[\protect\citeauthoryear{{Karovska}, {Gaetz}, {Carilli}, {Hack},
  {Raymond}  \& {Lee}}{{Karovska} et~al.}{2010}]{2010ApJ...710L.132K}
{Karovska} M.,  {Gaetz} T.~J.,  {Carilli} C.~L.,  {Hack} W.,  {Raymond} J.~C.,
   {Lee} N.~P.,  2010, \mn@doi [\apjl] {10.1088/2041-8205/710/2/L132}, \href
  {https://ui.adsabs.harvard.edu/abs/2010ApJ...710L.132K} {710, L132}

\bibitem[\protect\citeauthoryear{{Kashyap}, {van Dyk}, {Connors}, {Freeman},
  {Siemiginowska}, {Xu}  \& {Zezas}}{{Kashyap}
  et~al.}{2010}]{2010ApJ...719..900K}
{Kashyap} V.~L.,  {van Dyk} D.~A.,  {Connors} A.,  {Freeman} P.~E.,
  {Siemiginowska} A.,  {Xu} J.,   {Zezas} A.,  2010, \mn@doi [\apj]
  {10.1088/0004-637X/719/1/900}, \href
  {https://ui.adsabs.harvard.edu/abs/2010ApJ...719..900K} {719, 900}

\bibitem[\protect\citeauthoryear{{Kennea}, {Mukai}, {Sokoloski}, {Luna},
  {Tueller}, {Markwardt}  \& {Burrows}}{{Kennea}
  et~al.}{2009}]{2009ApJ...701.1992K}
{Kennea} J.~A.,  {Mukai} K.,  {Sokoloski} J.~L.,  {Luna} G.~J.~M.,  {Tueller}
  J.,  {Markwardt} C.~B.,   {Burrows} D.~N.,  2009, \mn@doi [\apj]
  {10.1088/0004-637X/701/2/1992}, \href
  {https://ui.adsabs.harvard.edu/abs/2009ApJ...701.1992K} {701, 1992}

\bibitem[\protect\citeauthoryear{Kolmogorov}{Kolmogorov}{1933}]{kolmogorov}
Kolmogorov A.,  1933, Giornale dell'Instituto Italiano degli Attuari, 4, 83

\bibitem[\protect\citeauthoryear{{Kraft} et~al.,}{{Kraft}
  et~al.}{2022}]{2022arXiv221109827K}
{Kraft} R.,  et~al., 2022, arXiv e-prints, \href
  {https://ui.adsabs.harvard.edu/abs/2022arXiv221109827K} {p. arXiv:2211.09827}

\bibitem[\protect\citeauthoryear{Kuehl}{Kuehl}{2000}]{kuehl2000}
Kuehl R.~O.,  2000, Designs of experiments: statistical principles of research
  design and analysis.
Duxbury press

\bibitem[\protect\citeauthoryear{Ledwina}{Ledwina}{1994}]{Led1994}
Ledwina T.,  1994, Journal of the American Statistical Association, 89, 1000

\bibitem[\protect\citeauthoryear{{Luna} \& {Sokoloski}}{{Luna} \&
  {Sokoloski}}{2007}]{2007ApJ...671..741L}
{Luna} G.~J.~M.,  {Sokoloski} J.~L.,  2007, \mn@doi [\apj] {10.1086/522576},
  \href {https://ui.adsabs.harvard.edu/abs/2007ApJ...671..741L} {671, 741}

\bibitem[\protect\citeauthoryear{{Luna} et~al.,}{{Luna}
  et~al.}{2018}]{2018A&A...616A..53L}
{Luna} G.~J.~M.,  et~al., 2018, \mn@doi [\aap] {10.1051/0004-6361/201832592},
  \href {https://ui.adsabs.harvard.edu/abs/2018A&A...616A..53L} {616, A53}

\bibitem[\protect\citeauthoryear{Miller}{Miller}{1977}]{m1977}
Miller R.~G.,  1977, Journal of the American Statistical Association, 72, 779

\bibitem[\protect\citeauthoryear{Moran}{Moran}{1973}]{moran1973}
Moran P. A.~P.,  1973, Sankhyā: The Indian Journal of Statistics, Series A,
  35, 329

\bibitem[\protect\citeauthoryear{{Muerset}, {Wolff}  \& {Jordan}}{{Muerset}
  et~al.}{1997}]{1997A&A...319..201M}
{Muerset} U.,  {Wolff} B.,   {Jordan} S.,  1997, \aap, \href
  {https://ui.adsabs.harvard.edu/abs/1997A&A...319..201M} {319, 201}

\bibitem[\protect\citeauthoryear{Mukhopadhyay}{Mukhopadhyay}{2017}]{deep2017}
Mukhopadhyay S.,  2017, Electronic Journal of Statistics, 11, 215

\bibitem[\protect\citeauthoryear{Neyman}{Neyman}{1937}]{neyman1937}
Neyman J.,  1937, Scandinavian Actuarial Journal., pp 149--199

\bibitem[\protect\citeauthoryear{Park, Van~Dyk  \& Siemiginowska}{Park
  et~al.}{2008}]{park}
Park T.,  Van~Dyk D.~A.,   Siemiginowska A.,  2008, The Astrophysical Journal,
  688, 807

\bibitem[\protect\citeauthoryear{Parzen}{Parzen}{2004}]{parzen2004}
Parzen E.,  2004, Statistical Science, 19, 652

\bibitem[\protect\citeauthoryear{Pearson}{Pearson}{1900}]{pearson}
Pearson K.,  1900, The London, Edinburgh, and Dublin Philosophical Magazine and
  Journal of Science, 50, 157

\bibitem[\protect\citeauthoryear{{Primini} \& {Kashyap}}{{Primini} \&
  {Kashyap}}{2014}]{2014ApJ...796...24P}
{Primini} F.~A.,  {Kashyap} V.~L.,  2014, \mn@doi [\apj]
  {10.1088/0004-637X/796/1/24}, \href
  {https://ui.adsabs.harvard.edu/abs/2014ApJ...796...24P} {796, 24}

\bibitem[\protect\citeauthoryear{{Reimers} \& {Cassatella}}{{Reimers} \&
  {Cassatella}}{1985}]{1985ApJ...297..275R}
{Reimers} D.,  {Cassatella} A.,  1985, \mn@doi [\apj] {10.1086/163525}, \href
  {https://ui.adsabs.harvard.edu/abs/1985ApJ...297..275R} {297, 275}

\bibitem[\protect\citeauthoryear{Smirnov}{Smirnov}{1939}]{smirnov}
Smirnov N.,  1939, Bull. Math. Univ. Moscou, 2, 3

\bibitem[\protect\citeauthoryear{{Smith}}{{Smith}}{2020}]{2020SPIE11444E..2CS}
{Smith} R.~K.,  2020, in Society of Photo-Optical Instrumentation Engineers
  (SPIE) Conference Series. p. 114442C, \mn@doi{10.1117/12.2576047}

\bibitem[\protect\citeauthoryear{{Smith}, {Brickhouse}, {Liedahl}  \&
  {Raymond}}{{Smith} et~al.}{2001}]{2001ApJ...556L..91S}
{Smith} R.~K.,  {Brickhouse} N.~S.,  {Liedahl} D.~A.,   {Raymond} J.~C.,  2001,
  \mn@doi [\apjl] {10.1086/322992}, \href
  {https://ui.adsabs.harvard.edu/abs/2001ApJ...556L..91S} {556, L91}

\bibitem[\protect\citeauthoryear{Wilks}{Wilks}{1938}]{wilks}
Wilks S.~S.,  1938, The annals of mathematical statistics, 9, 60

\bibitem[\protect\citeauthoryear{{XRISM Science Team}}{{XRISM Science
  Team}}{2022}]{2022arXiv220205399X}
{XRISM Science Team} 2022, arXiv e-prints, \href
  {https://ui.adsabs.harvard.edu/abs/2022arXiv220205399X} {p. arXiv:2202.05399}

\makeatother
\end{thebibliography}

\appendix

\section{Proof of the Convergence of K-statistic} \label{appendixA}

Let $k_m$ be the observed value of the K-statistic $K_m$ on the data and $W_j\stackrel{i.i.d}{\sim} \chi^2_1$ for $j=1,...,m$, we have
\begin{equation*}
    \begin{aligned}
P\Big(K_m > k_m \ | \ H_0\Big) &= 1 - P\Big(K_m\leq k_m \ | \ H_0 \Big) \\
&= 1- P\Big(\mathop{\max}_{j=1,..,m}n\widehat{\theta_j}^2 \leq k_m \ | \ H_0 \Big) \\
&\stackrel{d}{\rightarrow} 1- P\Big(\mathop{\max}_{j=1,..,m} W_j \leq  k_m \Big) \\
&= 1- P\Big(\chi^2_1\leq k_m \Big)^{m} \,,
    \end{aligned}    
\end{equation*}
corresponding to Equation~\eqref{eqn:kpval}, where the notation $\stackrel{d}{\rightarrow}$ denotes convergence in distribution, and the convergence is for $n\to\infty$.

\section{Proof of Theorem 1} \label{appendixB}
Consider the event $E_m$ defined as
\[E_m = \Big \{\text{$m$ maximises the BIC criterion in \eqref{eqn:BIC}} \Big \}\]
Then, an upper bound for the limit of $P(K_{(m)}\geq k_{(m)}\ | \ H_0)$ is
\begin{equation*}
    \begin{aligned}
P(K_{(m)}&\geq k_{(m)}\ | \ H_0) = P\Big (K_{m} > k_m \ \cap \  E_m \ |\ H_0 \Big)\\ &=  P \Big(\mathop{\max}_{j=1,..,m} n\widehat{\theta}_{(j)}^2 > k_m \ \cap \ E_m \ |\ H_0 \Big)\\
&= P\Big(\mathop{\max}_{j=1,..,m} n\widehat{\theta}_{(j)}^2 > k_m \ | E_m \cap \  H_0  \Big) P\Big(E_m \ | \ H_0 \Big)\\
&= P\Big(\mathop{\max}_{j=1,..,M} n\widehat{\theta}_{(j)}^2 > k_m\ | E_m \cap \  H_0  \Big) P\Big(E_m \ | \ H_0 \Big)\\ 
&\leq \sum_{m=1}^{M} P\Big(\mathop{\max}_{j=1,..,M} n\widehat{\theta}_{(j)}^2 > k_m \ | E_m \cap \  H_0  \Big) P \Big(E_m \ | \ H_0 \Big)\\ 
&= \ P\Big(\mathop{\max}_{j=1,..,M} n\widehat{\theta}_{(j)}^2 > k_m \ | \ H_0 
\Big)
\stackrel{d}{\rightarrow} 1- P\Big( \chi^2_1\leq k_m\Big)^M \\
    \end{aligned}    
\end{equation*}
where the convergence is for $n\rightarrow \infty$, and follows from (\ref{eqn:kpval}). This proves Theorem~1 (Equation~\eqref{eqn:kadjpval}).

\section{Corrected Background Density} \label{appendixD}
The corrected background density for region $\passband{4}$ (Equation~\eqref{eqn:corrected}) is calculated as:
\addtolength{\jot}{2pt}
\begin{equation*}
    \begin{aligned}
\widehat{b}_{W_4}(x)&= \frac{1}{\int_{l_4}^{u_4} \widehat{b}_{C_4}(x)} \widehat{b}_{C_4}(x)\\ 
&= \frac{1}{\int_{14.6}^{15.15} (1.3994 - 0.0250x)}(1.3994 - 0.0250x) \\
&= 2.4749 - 0.0441x 
    \end{aligned}
\end{equation*}
where $\widehat{b}_{C_4}$ is the corrected background density for the combined region $C_4$ in \eqref{eqn:corre}; $l_4$, $u_4$ corresponds to the lower and upper bounds of the wavelength range of region $\passband{4}$ and $x\in [14.6,15.15]$ \AA. The corrected background density for region $\passband{5}$ can be derived in a similar manner. 

\section{Additional tables}
Tables \ref{table:var-importance} and  \ref{table:var-importance_sidak} report the probability of Type I error for our smooth tests and adequate post-selection and multiple hypothesis testing adjustments. Specifically, Table \ref{table:var-importance} reports the type I errors inclusive only of the post-selection adjustments. Specifically, this corresponds to the probabilities of false discoveries we would expect our testing procedures to have if we were to conduct an analysis on each of the regions considered individually, that is, without simultaneously testing for spectral lines in the remaining six regions. Conversely, in Table \ref{table:var-importance_sidak}, we report the probability of a false discovery across the entire spectrum, that is, when accounting for the fact that seven different tests are being conducted simultaneously. 

Tables \ref{table:vsd} and \ref{table:vsd2} report the upper limits obtained using smooth tests for the spectral lines of interest. Similarly to the upper limits obtained by means of the LRT, Table \ref{table:vsd} and Table \ref{table:vsd2} provide the local and Sidak adjusted upper limits respectively described in detail in Section \ref{sec:constructUL}. 

\begin{table*}
    \centering
	\begin{tabular}{
	|c!{\vrule width 1.4pt}
	c|c|c|c!{\vrule width 1.4pt}}
		\hline
		Regions & Bonferroni & K & Naive &LRT\\
			of interest ($\passband{r}$)&   &   && \\
		\hline
	    $\passband{3}$& 0.0511 ($\pm$0.0022)& 0.0509 ($\pm$0.0022) & 2e-4 ($\pm$1e-4)  & 0.0434 ($\pm$0.0020)\\
		$\passband{4}$& 0.0559 ($\pm$0.0023)& 0.0500 ($\pm$0.0022) & 3e-4 ($\pm$2e-4)  &0.0445 ($\pm$0.0021)\\
		$\passband{5}$ & 0.0472 ($\pm$0.0021)& 0.0455 ($\pm$0.0021) & 3e-4  ($\pm$2e-4)  &0.0422 ($\pm$0.0020)\\
		$\passband{6}$& 0.0511 ($\pm$0.0022) & 0.0487 ($\pm$0.0022) & 3e-4  ($\pm$2e-4)  &0.0446 ($\pm$0.0021)\\
	$\passband{7}$ & 0.0602 ($\pm$0.0024)& 0.0497 ($\pm$0.0022)  & 1e-4 ($\pm$1e-4)  &0.0449 ($\pm$0.0021)\\
	$\passband{8}$  & 0.0612 ($\pm$0.0024)& 0.0474 ($\pm$0.0021)& 3e-4 	($\pm$2e-4) & 0.0471 ($\pm$0.0021)\\
	$\passband{9}$ & 0.0619 ($\pm$0.0024)& 0.0475 ($\pm$0.0021) & 3e-4  ($\pm$2e-4) & 0.0401 ($\pm$0.0020)\\
		\hline
	\end{tabular}
      \caption{Simulated type I error and respective Monte Carlo errors for the smooth tests computed on regions $\passband{3}$-$\passband{9}$ and setting the local significance level to $0.05$.}
	\label{table:var-importance}
\end{table*}

\begin{table*}
    \centering
	\begin{tabular}{
	|c!{\vrule width 1.4pt}
	c|c|c|c!{\vrule width 1.4pt}}
		\hline
		Regions & Bonferroni & K & Naive & LRT\\
			of interest ($\passband{r}$)&   &   & &\\
		\hline               
	    $\passband{3}$& 0.0118 ($\pm$0.0023)& 0.0082 ($\pm$0.0009) & 0 ($\pm$0)  & 0.0070 ($\pm$0.0008)\\
		$\passband{4}$& 0.0130 ($\pm$0.0031)& 0.0072 ($\pm$0.0008) & 0 ($\pm$0)  & 0.0057 ($\pm$0.0008)\\
		$\passband{5}$ & 0.0102 ($\pm$0.0030)& 0.0063 ($\pm$0.0008) & 0 ($\pm$0)  & 0.0063 ($\pm$0.0008)\\
		$\passband{6}$& 0.0114 ($\pm$0.0024) & 0.0065  ($\pm$0.0008) & 0 ($\pm$0) & 0.0058 ($\pm$0.0008)\\
	$\passband{7}$ &  0.0133 ($\pm$0.0033)&  0.0062 ($\pm$0.0008)  & 0 ($\pm$0)  &0.0060 ($\pm$0.0008)\\
	$\passband{8}$  & 0.0139 ($\pm$0.0034)&  0.0058 ($\pm$0.0008)& 0		($\pm$0) & 0.0070 ($\pm$0.0008)\\
	$\passband{9}$ & 0.0142 ($\pm$0.0029)& 0.0078 ($\pm$0.0009) & 0 ($\pm$0) & 0.0056 ($\pm$0.0007)\\
		\hline
	\end{tabular}
      \caption{Simulated type I error and respective Monte Carlo errors for the smooth tests computed on regions $\passband{3}$-$\passband{9}$ and adjusted for multiple hypothesis testing using Sidak's correction with global significance $\alpha=0.05$. In this case, the local significance is $\alpha_r=1-(1-\alpha)^{\frac{1}{R}} = 0.0073$, for $r=1,\dots,R$.}
	\label{table:var-importance_sidak}
\end{table*}

\begin{table*}
	\centering
	\begin{tabular}{
	|c!{\vrule width 1.4pt}
	ccc!{\vrule width 1.4pt}
	ccc!{\vrule width 1.4pt}}
		\hline
		Local& 
		\multicolumn{3}{c!{\vrule width 1.4pt}}{50$\%$ Upper limits via smooth tests} & \multicolumn{3}{c!{\vrule width 1.4pt}}{90$\%$ Upper limits via smooth tests } \\
		\cline{1-7}
Regions ($\passband{r}$)	& Bonferroni & K & Naive & Bonferroni& K & Naive\\
		\hline 
	    $\passband{3}$&45.26 & 45.99 & 61.32 & 70.08 & 71.54 & 85.41\\
		$\passband{4}$ &28.90 & 31.37 & 38.53 & 44.71 &48.17 &53.60 \\
		$\passband{5}$&35.32 &36.74 &47.57 & 55.11 &56.99 &65.47\\
		$\passband{6}$&32.76  & 34.32 &  44.07 & 51.09 &53.82 &60.84 \\
		$\passband{7}$ &26.85 &30.07 &36.70 & 42.06 &45.82 &50.12   \\
		$\passband{8}$ &26.82 &29.87 &37.56&41.76 &45.53 &50.75\\
		$\passband{9}$ & 58.50 &66.30 &79.56& 91.65 &100.6 &110.0 \\
		\hline
	\end{tabular}
      \caption{$50\%$ and $90\%$ upper limits obtained using smooth tests with different post-selection adjustments. The $50\%$ upper limits are calculated by the proportion of the expected lines $\eta$ to achieve the power 0.5 (solid line in Figure \ref{fig:compare_fig}) times the sample sizes $n_r$. Similarly, the $90\%$ upper limits are calculated by the proportion of the expected lines $\eta$ to achieve the power 0.9 times the sample sizes $n$.}
	\label{table:vsd}
\end{table*}
\begin{table*}
	\centering
	\begin{tabular}{
	|c!{\vrule width 1.4pt}
	ccc!{\vrule width 1.4pt}
	ccc!{\vrule width 1.4pt}}
		\hline
	Sidak adjusted& 
		\multicolumn{3}{c!{\vrule width 1.4pt}}{50$\%$ upper limits via smooth tests} & \multicolumn{3}{c!{\vrule width 1.4pt}}{90$\%$ upper limits via smooth tests}\\
		\cline{1-7}
Regions ($\passband{r}$)	&Bonferroni &K&Naive&Bonferroni &K&Naive\\
	\hline  
	    $\passband{3}$&53.29 & 63.88 & 67.16  & 77.38 & 89.06&89.06\\
		$\passband{4}$  &34.09 & 42.48 & 42.73& 49.40& 59.77&57.55\\
		$\passband{5}$&41.45 &50.40& 52.28 & 58.88 &69.24&70.18\\
		$\passband{6}$&39.00 &47.19 &48.75& 56.16 &66.69&65.52\\
		$\passband{7}$ &32.76 &41.35 &41.17  & 46.54 &56.74&53.70\\
		$\passband{8}$ &33.20& 39.88 &42.20 &46.55 &54.38 &55.68 \\
		$\passband{9}$ & 71.37 & 91.26& 89.31& 101.4& 125.2 &118.2 \\
		\hline
	\end{tabular}
      \caption{Sidak's corrected $50\%$ and $90\%$  upper limits obtained using smooth tests with different post-selection adjustments.}
	\label{table:vsd2}
\end{table*}

\label{lastpage}
\end{document}